\documentclass[aps,prc,reprint,showpacs,superscriptaddress,onecolumn,notitlepage,nofootinbib]{revtex4-1}

\usepackage{graphicx,color} 
\usepackage{bm}       
\usepackage{amsmath}
\usepackage[dvipsnames]{xcolor}
\usepackage{ulem}
\usepackage{siunitx}
\usepackage{multirow}
\usepackage{gensymb}
\usepackage{dcolumn}

\begin{document}

\newcommand {\nc} {\newcommand}
\nc {\IR} [1]{\textcolor{red}{#1}}
\nc {\IB} [1]{\textcolor{blue}{#1}}
\nc {\IP} [1]{\textcolor{magenta}{#1}}
\nc {\IM} [1]{\textcolor{Bittersweet}{#1}}
\nc {\IE} [1]{\textcolor{Plum}{#1}}

\title{Chiral uncertainties in \textit{ab initio} nucleon-nucleus elastic scattering }

\author{R.~B.~Baker}
\affiliation{Institute of Nuclear and Particle Physics, and Department of Physics and Astronomy, Ohio
University, Athens, OH 45701, USA}
\author{M.~Burrows}
\affiliation{Department of Physics and Astronomy, Louisiana State University, Baton Rouge, LA 70803,
USA}
\author{Ch.~Elster}
\affiliation{Institute of Nuclear and Particle Physics, and Department of Physics and Astronomy, Ohio
University, Athens, OH 45701, USA}
\author{K.D. Launey}
\affiliation{Department of Physics and Astronomy, Louisiana State University, Baton Rouge, LA 70803,
USA}
\author{P.~Maris}
\affiliation{Department of Physics and Astronomy, Iowa State University, Ames, IA 50011, USA}
\author{G.~Popa}
\affiliation{Institute of Nuclear and Particle Physics, and Department of Physics and Astronomy, Ohio
University, Athens, OH 45701, USA}
\author{S.~P.~Weppner}
\affiliation{Natural Sciences, Eckerd College, St. Petersburg, FL 33711, USA}

\begin{abstract}

The effective interaction between a nucleon and a nucleus is one of the most important ingredients
for reaction theories. Theoretical formulations were introduced early by Feshbach and Watson, and
efforts of deriving and computing those `optical potentials'  in a microscopic fashion
have a long tradition.  However, only recently the leading order term in the Watson multiple
scattering
approach could be calculated fully {\it ab initio}, meaning that the same nucleon-nucleon (NN)
interaction enters
both the structure as well as the reaction pieces on equal footing. This allows the
uncertainties from the underlying chiral effective NN interaction to be systematically explored in
nucleon-nucleus elastic scattering observables.

In this contribution the main ingredients for arriving at the {\it ab initio} leading order of the
effective
nucleon-nucleus interaction in the Watson approach will be reviewed. Concentrating on one
specific chiral NN interaction from the LENPIC collaboration and light nuclei with a 0$^+$ ground
state,
the leading order nucleon-nucleus interaction is calculated using up to the third chiral order (N2LO) in the
nucleon-nucleon potential, and
elastic scattering observables are extracted. Then pointwise as well as correlated uncertainty
quantification is used for the estimation of the chiral truncation error. Elastic scattering
observables for $^4$He, $^{12}$C, and $^{16}$O for between 65 and 200 MeV projectile energy will be
analyzed.

\end{abstract}


\maketitle

\section{Introduction}

Simplifying the many-body problem posed by scattering of a proton or neutron from a nucleus to a
two-body problem with an effective (optical) potential was introduced already by Bethe~\cite{Bethe:1935zz}
in the 1930s, and its justification summarized by Feshbach~\cite{Feshbach:1958wf}. 
Since then differential cross sections as well as spin observables for elastic
scattering played an important role in either determining the parameters in phenomenological optical
models for proton or neutron scattering from nuclei or in testing validity and accuracy of
microscopic models thereof.  The theoretical approach to elastic scattering from a nuclear target 
presented in this article is based on the ansatz of a multiple scattering expansion that 
was pioneered by
Watson~\cite{Watson1953a,Watson1953b}, made familiar by Kerman, McManus, and Thaler (KMT)~\cite{KMT}.
 and refined further as spectator expansion~\cite{Siciliano:1977zz,Ernst:1977gb,Tandy:1980zz}. Specifically, elastic scattering
from stable nuclei has led in the 1990s to a large body of work on microscopic optical
potentials in which the nucleon-nucleon interaction and the density of the nucleus were taken as
input to rigorous calculations of first-order potentials, in either a Kerman-McManus-Thaler (KMT) or
a Watson expansion of the multiple scattering series
(see
e.g.~\cite{Crespo:1992zz,Crespo:1990zzb,Elster:1996xh,Elster:1989en,Arellano:1990xu,Arellano:1990zz}).
Here the primary goal was a deeper understanding of the reaction mechanism.
However, a main disadvantage of that work was the lack of sophisticated nuclear structure input
compared to what is available today.  

Recent developments of the nucleon-nucleon (NN) and three-nucleon (3N) interactions, derived
from chiral effective field theory, have yielded major
progress~\cite{EntemM03,Epelbaum06,Epelbaum:2008ga,Epelbaum:2014sza,Epelbaum:2014efa,Reinert:2017usi,Machleidt:2011zz,Entem:2017gor}.
These, together with the utilization of massively parallel computing resources (e.g.,
see~\cite{LangrDDLT19,LangrDDT18,SHAO20181,CPE:CPE3129,Jung:2013:EFO}), have placed {\it ab initio}
large-scale simulations at the frontier of nuclear structure and reaction explorations. Among other
successful many-body theories, the {\it ab initio} no-core shell-model (NCSM) approach (see, e.g.,
\cite{Navratil:2000ww,Roth:2007sv,BarrettNV13,Binder:2018pgl}), has over
the last decade taken center stage in the development of microscopic tools for studying the
structure of atomic nuclei. The NCSM
concept combined with a symmetry-adapted (SA) basis in the {\it ab initio} SA-NCSM
\cite{LauneyDD16} has further expanded the reach to the structure of intermediate-mass
nuclei~\cite{Dytrych:2020vkl}.
 
Following the developments in nuclear structure theory, it is very natural to again consider rigorous
calculations of effective folding nucleon-nucleus (NA) potentials, since now the nuclear densities
required as input for the folding with the NN scattering amplitudes can be based on the same chiral NN
interaction. This development also allows to investigate effects of truncation uncertainties  in 
the chiral expansion on NA scattering observables in a similar fashion as already successfully
performed in NN scattering (see e.g.~\cite{Furnstahl:2015rha,Melendez:2017phj,Melendez:2019izc}),
nucleon-deuteron scattering~\cite{Epelbaum:2019zqc}, or structure observables for light
nuclei~\cite{Binder:2018pgl,Maris:2020qne}.

The theoretical and computational developments leading to {\it ab initio} NA effective interactions
(in leading order in the spectator expansion) are described in a serious of
publications by the authors~\cite{Burrows:2017wqn,Burrows:2018ggt,Burrows:2020qvu,Baker:2021izp,Baker:2021iqy}
and others~(see e.g.~\cite{Gennari:2017yez,Vorabbi:2021kho,Vorabbi:2020cgf,Arellano:2022tsi}). Thus
the aim of this review is to shed light on truncation uncertainties in the chiral expansion,
and within that context give a perspective on intricacies of the spectator expansion as well as the explicit content
of its leading order term, which can now be calculated {\it ab initio}.

Deriving {\it ab initio} optical potentials within a multiple scattering approach focuses on projectile energies at energies about 80~MeV or higher, since the expectation is that at those energies the leading order term may already capture the most important physics. Another recent {\it ab initio} approach starts from a formulation introduced by Feshbach~\cite{Feshbach:1958nx} and constructs optical potentials and elastic scattering observables within a Green's function approach~\cite{Dickhoff:2004xx,Idini:2019hkq}. For elastic scattering from medium-mass nuclei the coupled-cluster method~\cite{Rotureau:2016jpf} and the SA-NCSM~\cite{Launey:2021sua} approach 
have been successfully implemented. These approaches are by design better suited for calculating scattering observables at energies below about 20-30~MeV due to restrictions on the size of the model spaces which increase with increasing projectile energy.
In Ref.~\cite{Hebborn:2022vzm} an extensive overview of the status of the field of optical potentials and their need in the rare-isotope era is given and the current status of {\it ab initio} approaches is discussed. We want to encourage the reader to refer to this work, for more details.
 

\section{Watson Optical Potential within the Spectator Expansion  }

The standard starting point for describing elastic scattering of a single projectile from a target
of $A$ particles  within a multiple scattering approach is the separation of the Lippmann-Schwinger
(LS) equation for the transition operator $T$,
\begin{equation}
T = V + V G_0 (E) T 
\label{eq:2.1}
\end{equation} 
into two parts, namely an integral equation for $T$,
\begin{equation}
T = U +U G_0 (E) P T,
\label{eq:2.2}
\end{equation}
where U is the effective potential operator defined by a second integral equation,
\begin{equation}
U = V +V G_0 (E) Q U.
\label{eq:2.3}
\end{equation}
Here $P$ is a projection onto the ground state of the target, $P=\frac{\left| \Phi_0
\right\rangle \left\langle \Phi_0 \right|}{\left\langle \Phi_0  | \Phi_0
\right\rangle}$, with $P+Q=1$ and $[G_0(E),P]=1$. The free propagator for the projectile and
target system is given by $G_0(E)=\left( E - h_0 - H_A + i\epsilon \right)^{-1}$ where $h_0$
is the kinetic energy of the projectile and $H_A$ is the Hamiltonian of the target nucleus. 
The general solutions of the nuclear  bound state problem $H_A |\Phi\rangle$ include the ground
state, excited states and continuum states. For the scattering problem given by the transition
amplitude $T$ the reference energy separating bound and continuum states is chosen such 
that the ground state energy is set to zero.
Thus energies referring to the target Hamiltonian in $G_0$ are excitation energies 
of the target. 
With these definitions the transition operator for elastic scattering may be redefined as $T_{\rm el}
= PTP$, in which case Eq.~(\ref{eq:2.2}) can be written as
\begin{equation}
T_{\rm el} = PUP +PUPG_0(E) T_{\rm el}.
\label{eq:2.4}
\end{equation}

\subsection{Spectator expansion of the operator $U$}
The transition operator for elastic scattering is given by a straightforward one-body integral
equation, which of course requires the knowledge of $PUP$, which is a many-body operator. 
For a brief review we follow the spectator expansion of $PUP$ as introduced in Ref.~\cite{Chinn:1995qn}
in contrast to Ref.~\cite{Siciliano:1977zz} where the expansion of $T$ is considered.  
Following those references, we assume the presence of two-body forces only for the present discussion. The
extension to many-body forces is not precluded by the formulation. 
With this assumption the operator $U$ can be expanded as 
\begin{equation}
U = \sum_{i=1}^A U_i,  \label{eq:2.5}
\end{equation}
where $U_i$ is given by
\begin{equation}
U_i = v_{0i} + v_{0i} G_0(E) Q \sum_{j=1}^A U_j ,  \label{eq:2.6}
\end{equation}
provided that $V= \sum_{i=1}^A v_{0i}$, where the two-body potential $v_{0i}$ acts between the
projectile and the $i$th target nucleon. Through the introduction of an operator $\tau_i$ which
satisfies 
\begin{equation}
\tau_i = v_{0i} + v_{0i} G_0(E) Q \tau_i , \label{eq:2.7}
\end{equation}
Eq.~(\ref{eq:2.6}) can be rearranged as
\begin{equation}
U_i = \tau_i + \tau_i G_0(E) Q \sum_{j \neq i} U_j . \label{eq:2.8}
\end{equation}
This rearrangement process can be continued for all $A$ target particles,
so that the operator for the optical potential can be expanded in  a series
of $A$ terms of the form
\begin{equation}
U = \sum_{i=1}^A \tau_i + \sum_{i,j \neq i}^A \tau_{ij} +
    \sum_{i,j \neq i,k \neq i,j}^A \tau_{ijk} + \cdots . \label{eq:2.9}
\end{equation}
This is the Spectator Expansion for $U$ , where each term is treated in turn.
The separation of the interactions according to the number of
interacting nucleons has a certain latitude, due to the
many-body nature of $G_0(E)$, which needs to be considered separately. In the following we will
concentrate on the leading-order term, which is still a many-body operator due the the presence of
$G_0(E)$. The next-to-leading order term in this spectator expansion for $U$ has been formally
derived and connected to standard three-body equations in Ref.~\cite{Chinn:1995qn}.

\subsection{Propagator expansion in the leading-order term of $U$ }
When using the leading-order term of the spectator expansion as given in Eq.~(\ref{eq:2.7}), for
elastic scattering only $P\tau_i P$, or equivalently
$\langle\Phi_0| \tau_i | \Phi_0\rangle$ needs be considered. With this in mind, Eq.~(\ref{eq:2.7})
can be re-expressed as \begin{eqnarray}
\tau_i = v_{0i} + v_{0i}G_0(E) \tau_i - v_{0i} G_0(E) P \tau_i
   = \hat{\tau_i} - \hat{\tau_i} G_0(E) P \tau_i, \label{eq:2.10}
\end{eqnarray}
or
\begin{equation}
\langle\Phi_0| \tau_i | \Phi_0\rangle = \langle\Phi_0| \hat{\tau_i}|
 \Phi_0\rangle - \langle\Phi_0| \hat{\tau_i}| \Phi_0\rangle \frac {1}
  {(E-E_A) - h_0 + i\varepsilon} \langle\Phi_0| \tau_i | \Phi_0\rangle,
 \label{eq:2.11}
\end{equation}
where $\hat{\tau_i}$ is defined as the solution of
\begin{equation}
\hat{\tau_i} = v_{0i} + v_{0i} G_0(E) \hat{\tau_i}. \label{eq:2.12}
\end{equation}

The combination of Eqs.~(\ref{eq:2.10}) and (\ref{eq:2.2}) corresponds
to the leading-order Watson optical potential \cite{Watson1953a,Watson1953b}. In {\it ab initio} structure
calculations the one-body densities or ground state wave functions
 for protons and neutrons are calculated separately, so that
Eq.~(\ref{eq:2.11}) allows to combine e.g. for proton scattering of a nucleus the proton-neutron interaction
(${\hat \tau}_{i=pn}$)  
 with the neutron one-body density and the proton-proton interaction with the proton one-body density.
The sum over $i$ then adds both to obtain the driving term $\langle\Phi_0| \hat{\tau_i}|\Phi_0\rangle$ the
integral equation, Eq.~(\ref{eq:2.11}).

\begin{figure}
\begin{center}
\includegraphics[width=0.55\textwidth]{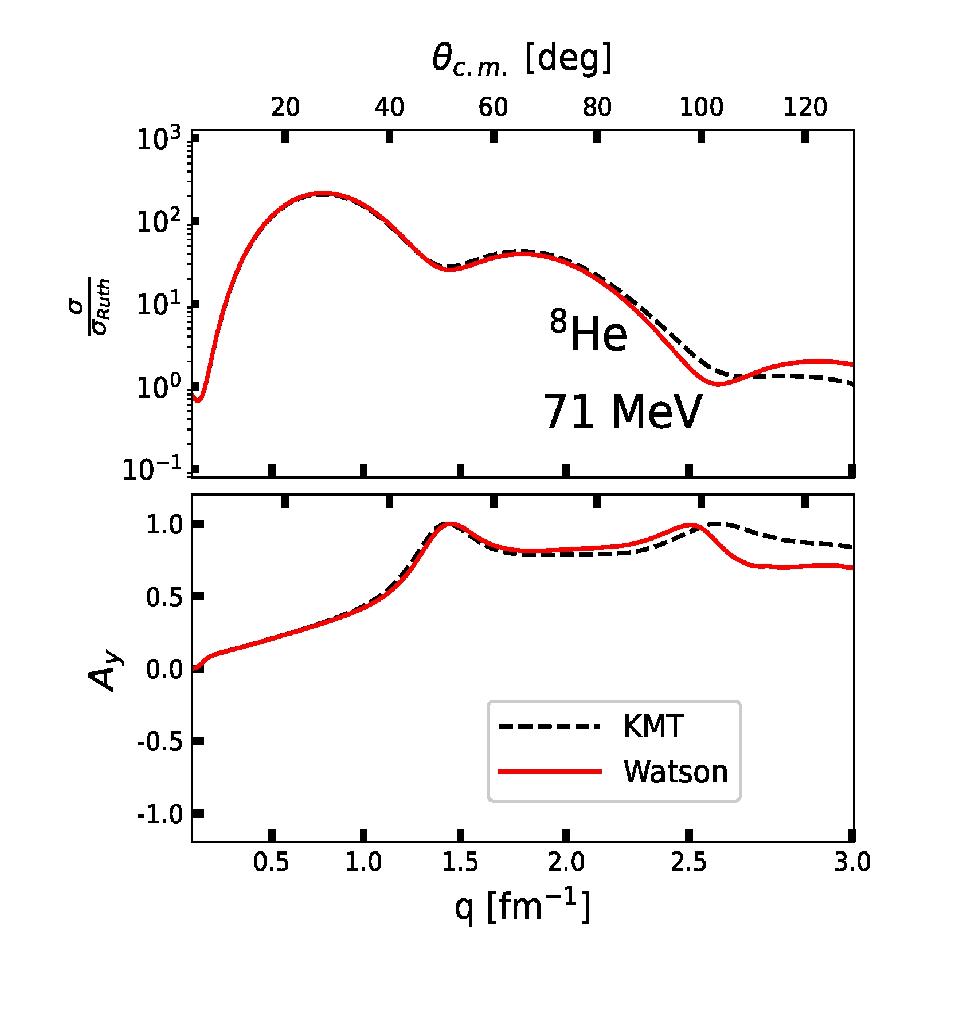}
\end{center}
\caption{The angular distribution of the differential cross section divided by the Rutherford cross
section (upper panel) and the analyzing power ($A_y$) for elastic proton scattering from $^8$He at 
71~MeV laboratory kinetic energy as function of the
momentum transfer $q$ and the c.~m. angle calculated with the LENPIC SCS chiral interaction~\cite{Epelbaum:2014efa}
with a cutoff $R=1$ fm. The
calculations are based on nonlocal densities using $\hbar\Omega = 14$ MeV at $N_{\rm max}=14$. The
solid (red) line stands for using the Watson optical potential while the black (dashed) line
represents the KMT prescription.
}
\label{fig0}
\end{figure}

 If the projectile-target-nucleon interaction is assumed to be the
same for all target nucleons and if iso-spin effects are neglected
then the KMT approximation ($\frac{A-1}{A}\langle\Phi_0| \hat{\tau_i}|\Phi_0\rangle)$
  can be derived from
the leading-order Watson potential~\cite{KMT}. When working with momentum space integral equations, the
numerical implementation of Eq.~(\ref{eq:2.11}) is
straightforward~\cite{Chinn:1993zza,Burrows:2018ggt,Burrows:2020qvu,Vorabbi:2021kho}. Working in
coordinate space with differential equations does not allow an equally straightforward implementation,
and thus the KMT prescription is the most favorable alternative. 
A comparison
between leading-order Watson potential and the KMT prescription is shown in Fig.~\ref{fig0} 
for elastic proton scattering
from $^8$He at 71~MeV laboratory kinetic energy. Despite the relatively large difference between the
proton and neutron densities for this nucleus the KMT prescription agrees with the exact Watson
description very well up to  momentum transfers of about 2~fm$^{-1}$.

Since Eq.~(\ref{eq:2.11}) is a  one-body integral equation,
the principal problem is to find a solution of Eq.~(\ref{eq:2.12}), which due to many-body character
of $G_0(E)$ is still a many-body integral equation, and in fact no more easily solved than the
starting point of Eq.~(\ref{eq:2.1}).

For most practical calculations the so-called  closure approximation to $G_0(E)$ is
implemented~\cite{Miller:1978zz}
turning Eq.~(\ref{eq:2.12}) into
a one-body integral equation. This approximation replaces $H_A$ by a constant that is interpreted
as an average excitation energy,  
and is justified when the projectile energy is large
compared to typical excitation energies of the nucleus. The closure approximation 
is very successfully applied for elastic scattering around 80~MeV and higher. 

Going beyond the closure approximation 
in the spirit of the spectator expansion we want to single out one target nucleon $i$ and write
 $G_0(E)$ as
\begin{eqnarray}
G_0(E) &=& ( E - h_0 -H_A + i\varepsilon)^{-1}  \nonumber \\
    &=& ( E - h_0 -h_i - \sum_{j\neq i} v_{ij} -H^i + i\varepsilon)^{-1}, \label{eq:2.13}
\end{eqnarray}
where the target Hamiltonian is expanded as $H_A = h_i +\sum_{j\neq i} v_{ij} +H^i$ with
$v_{ij}$ being the interaction between target nucleons $i$ and $j$, and $H^i$ being an (A-1)-body
operator containing all higher order effects.
Realizing that $\sum_{j\neq i} v_{ij} \equiv W_i$ and thus $H^i = H_A -h_i -W_i$ does not have an explicit
dependence on the $i$th particle, then $H^i$ may be replaced by an average energy $E^i$ 
which is akin to the effective binding energy between the $i$th nucleon and the $A-1$ spectator.
This is not an approximation since $G_0 (E)$ may be regarded as 
\begin{equation}
G_0(E) = [(E-E^i) -h_0 -h_i -W_i - (H^i -E^i) + i\varepsilon]^{-1}
 \label{eq:2.14}
\end{equation}
and $(H^i -E^i)$ should be set aside to be treated in the next order
of the expansion of the propagator $G_0(E)$. In this order of the expansion $G_0(E)$ becomes
\begin{equation}
G_i(E)= [(E-E^i) -h_0 -h_i -W_i + i\varepsilon]^{-1}, \label{eq:2.15}
\end{equation} 
and Eq.~(\ref{eq:2.12}) reads
\begin{equation}
\hat{\tau_i}=v_{0i} + v_{0i} G_i(E)\hat{\tau_i}. \label{eq:2.16}
\end{equation}
In order to connect the above expression with the free NN amplitude  
\begin{equation}
t_{0i} = v_{0i} + v_{0i} g_i t_{0i} \label{eq:2.17}
\end{equation}
with 
\begin{equation}
g_i =[ (E-E^i) - h_0 -h_i + i\varepsilon]^{-1}. 
\label{eq:2.18}
\end{equation}
algebraic relations between the resolvents lead to
\begin{equation}
\hat{\tau_i} = t_{0i} + t_{0i} G_i W_i g_i(E) \hat{\tau_i}.
                         \label{eq:2.19}
\end{equation}
Defining $G_i W_i = g_i {\mathcal T}_i$ with ${\mathcal T}_i = W_i +W_i g_i {\mathcal T}_i$ leads to
\begin{equation}
\hat{\tau_i} = t_{0i} + t_{0i} g_i {\mathcal T}_i g_i \hat{\tau_i}. \label{eq:2.20} 
\end{equation}
The three-body character of the above expression becomes more evident if one defines it as a set of
coupled equations as
\begin{eqnarray} 
\hat{\tau_i}&=& t_{0i} + t_{0i} g_i  X_i \nonumber \\
 X_i &=& {\mathcal T}_i g_i \hat{\tau_i}. \label{eq:2.21}
\end{eqnarray}

Though the spectator expansion of the operator $U$ in terms of active particles is defined in
Eq.~(\ref{eq:2.9}), we see that this expansion is performed in terms of quantities which 
contain many-body propagators. Each of the ingredients $\tau_i$, $\tau_{ij}$, etc. may themselves be
expanded in a spectator expansion, i.e. expanding the many-body propagator also according to the
number of active participants. The corrections to the propagator in the leading-order term of $U$
contributions that arise from the $Q$ space, whereas the terms arising from the propagator remain in
the $P$ space at first order level. Thus their contribution may be more relevant for elastic
scattering. 

In an explicit treatment of $G_i(E)$ it is necessary to consider the explicit form of $\sum_{j\neq
i} v_{ij} = W_i$, which is a priori a two-body operator. In the framework of {\it ab initio} nuclear
structure calculations this will involve two-body densities. In earlier
work~\cite{Chinn:1995qn,Chinn:1993zz,Chinn:1994xz} the quantity $W_i$ was treated as one-body
operator, specifically a mean-field potential. This was a physically reasonable choice, though being
outside the strict demands of the spectator expansion. However, those studies revealed that the next
order in the propagator expansion has little effect on elastic scattering observables at energies
larger than 100~MeV, while the description of differential cross section and spin-observables 
for elastic scattering from $^{40}$Ca at 48~MeV showed considerable improvement with respect to
experiment~\cite{Chinn:1993zz}. Obviously this type of calculation will need to be explored within an
{\it ab initio} approach.
In Ref.~\cite{Chinn:1993zz} the energy  $E^i$ of Eq.~(\ref{eq:2.18}) was set to zero.

As illustrated in this section, deriving a multiple scattering expansion for elastic NA scattering means
projecting on the ground state of the target in order to obtain a Lippman-Schwinger type equation for the
transition amplitude and obtaining an operator $U$ for the effective interaction, which is defined in the space
$Q=1-P$. In this spirit, the spectator expansion contains therefore two pieces, namely the expansion of the
operator $U$ in terms of active particles in the scattering process as well as the expansion of target 
Hamiltonian $H_A$ in the propagator $G_0(E)$ in a similar fashion. Thus it is very difficult to define a
single expansion parameter which governs the convergence of the expansion.


\section{Leading order {\textit {ab initio}} optical potential based on a chiral NN interaction}

The leading order of the spectator expansion involves two active nucleons, the projectile and a
target nucleon. Therefore, the leading order is driven by the NN amplitude $\overline M$ , which in its
most general form can be parameterized in terms of Wolfenstein
amplitudes~\cite{wolfenstein-ashkin,Fachruddin:2000wv,Golak:2010wz},
\begin{eqnarray}
\label{eq:3.1}
       \overline{M}(\bm{q},\bm{\mathcal{K}}_{NN},\epsilon)&=&
A(\bm{q},\bm{\mathcal{K}}_{NN},\epsilon)\textbf{1}\otimes\textbf{1} \nonumber \\
       &+& iC(\bm{q},\bm{\mathcal{K}}_{NN},\epsilon)~\left(\bm{\sigma^{(0)}} \cdot \hat{\bm{n}}
\right)\otimes\textbf{1} \nonumber \\
       &+& iC(\bm{q},\bm{\mathcal{K}}_{NN},\epsilon)~\textbf{1}\otimes\left(\bm{\sigma^{(i)}}\cdot
\hat{\bm{n}} \right) \nonumber \\
       &+&
M(\bm{q},\bm{\mathcal{K}}_{NN},\epsilon)(\bm{\sigma^{(0)}}\cdot\hat{\bm{n}})\otimes(\bm{\sigma^{(i)}}\cdot\hat{\bm{n}})
\nonumber \\
       &+&
\left[G(\bm{q},\bm{\mathcal{K}}_{NN},\epsilon)-H(\bm{q},\bm{\mathcal{K}}_{NN},\epsilon)\right](\bm{\sigma^{(0)}}\cdot\hat{\bm{q}})\otimes(\bm{\sigma^{(i)}}\cdot\hat{\bm{q}})
\cr
       &+&
\left[G(\bm{q},\bm{\mathcal{K}}_{NN},\epsilon)+H(\bm{q},\bm{\mathcal{K}}_{NN},\epsilon)\right](\bm{\sigma^{(0)}}\cdot\hat{\bm{\mathcal{K}}})\otimes(\bm{\sigma^{(i)}}\cdot\hat{\bm{\mathcal{K}}})
\cr
       &+&
D(\bm{q},\bm{\mathcal{K}}_{NN},\epsilon)\left[(\bm{\sigma^{(0)}}\cdot\hat{\bm{q}})\otimes(\bm{\sigma^{(i)}}\cdot\hat{\bm{\mathcal{K}}})+(\bm{\sigma^{(0)}}\cdot\hat{\bm{\mathcal{K}}})\otimes(\bm{\sigma^{(i)}}\cdot\hat{\bm{q}})\right]~,
\end{eqnarray}
where $\bm{\sigma^{(0)}}$ describes the spin of the projectile, and $\bm{\sigma^{(i)}}$ the spin
of the struck nucleon.
The average momentum in the NN frame is defined as $\bm{\mathcal{K}}_{NN} =
\frac{1}{2} \left(\bm{k'}_{NN} + \bm{k}_{NN}\right)$. The scalar functions $A$, $C$, $M$, $G$, $H$,
and $D$ are referred to as
Wolfenstein amplitudes and only depend on the scattering momenta and energy. Each term in
Eq.~(\ref{eq:3.1}) has two components, namely a scalar function of two vector momenta and an energy and the coupling between the operators
of the projectile and the struck nucleon.  The linear independent unit
vectors $\hat{\bm{q}}$, $\hat{\bm{\mathcal{K}}}$, and $\hat{\bm{n}}$ are defined in
terms of the momentum transfer and the average momentum as
\begin{eqnarray}
\hat{\bm{q}}=\frac{\bm{q}}{\left| \bm{q} \right|}~~,~~~
\hat{\bm{\mathcal{K}}}=\frac{\bm{\mathcal{K}}}{\left| \bm{\mathcal{K}} \right|} ~~,~~~
 \hat{\bm{n}}=\frac{\bm{\mathcal{K}} \times \bm{q}}{\left| \bm{\mathcal{K}} \times \bm{q}
\right|},
\label{eq:3.2}
\end{eqnarray}
and span the momentum vector space. With the exception of the momentum transfer $\bm q$, which is
invariant under frame transformation, the vectors in Eq.~(\ref{eq:3.2}) need to be considered in their
respective frame in explicit calculations~\cite{Burrows:2020qvu,BurrowsM:2020}. For the struck target
nucleon the expectation values of the operator ${\bf 1}$ and the scalar products of
$\bm{\sigma^{(i)}}$ with the linear independent unit vectors of Eq.~(\ref{eq:3.2}) need to be evaluated
with the ground state wave functions of the respective nucleus when calculating the leading-order NA
effective interaction. 
Evaluating the expectation value of the operator ${\bf 1}$ in the ground state of
the nucleus results in the scalar nonlocal, translationally invariant one-body density 
that has traditionally been used as input
to microscopic or {\it ab initio} calculations of leading order effective
interactions~\cite{Elster:1996xh,Elster:1989en,Burrows:2018ggt,Gennari:2017yez}.
The other operators from Eq.~(\ref{eq:3.2}), namely $(\bm{\sigma^{(i)}} \cdot \hat{\bm n})$,
$(\bm{\sigma^{(i)}} \cdot \hat{\bm q})$, and $(\bm{\sigma^{(i)}} \cdot \hat{\bm {\mathcal{K}}})$
need to also be evaluated for a leading-order {\it ab initio} NA effective interaction, in which the
NN interaction is treated on equal footing in the reaction and structure calculation.

Thus, the general expression for a nonlocal density needs to include the spin operator
$\bm{\sigma^{(i)}}$ explicitly,
\begin{equation}
\label{density}
\rho^{K_s}_{q_s}\left(\bm{p}, \bm{p}' \right) = \left\langle \Phi_0' \left| \sum_{i=1}^{A} \delta^3(
\bm{p_i} - \bm{p}) \delta^3( \bm{p_i}' - \bm{p}') \sigma_{q_s}^{(i) K_s} \right| \Phi_0 \right\rangle~,
\end{equation}
where $\sigma^{(i) K_s}_{q_s}$ is the spherical representation of the spin operator and the
wavefunction $\Phi_0$ $(\bm{p_1}, ..., \bm{p_A}) = \left\langle \bm{p_1}, ..., \bm{p_A} | \Phi_0
\right\rangle$ is defined in momentum space. Evaluating this expression for $K_s=0$ gives the nonlocal
one-body scalar density and $K_s=1$ becomes a nonlocal one-body spin density. 

The Wolfenstein parameterization of Eq.~(\ref{eq:3.1}) requires the evaluation of scalar products 
of the one-body spin density with unit momentum vectors. Since those only depend on the momenta
$\bm{p}$ and $\bm{p}'$, those can be calculated as 
$\rho^{K_s} \left(\bm{p}, \bm{p}' \right) \cdot \hat{\bm{n}}$, 
$\rho^{K_s} \left(\bm{p}, \bm{p}' \right) \cdot \hat{\bm{q}}$, and $\rho^{K_s}
\left(\bm{p}, \bm{p}' \right) \cdot \hat{\bm{\mathcal{K}}}$. 
For the explicit calculation of  $\rho^{K_s} \left(\bm{p}, \bm{p}' \right) \cdot
\hat{\bm{n}}$, we refer the reader to~\cite{Burrows:2020qvu,BurrowsM:2020}. The scalar products
$(\bm{\sigma^{(i)}} \cdot
\hat{\bm q})$ and $(\bm{\sigma^{(i)}} \cdot \hat{\bm {\mathcal{K}}})$ represent scalar products of a
pseudo-vector and a vector, a construct that is not invariant under parity transformations, and thus
vanish when sandwiched between ground state wave functions, which is explicitly shown
in~\cite{BurrowsM:2020}. Thus the tensor contributions of the NN force only enter the leading
order effective NA interaction through the Wolfenstein amplitude $M$ as long as elastic scattering is
considered. When e.g. transition amplitudes between states of different parity would be considered,
the other tensor amplitudes will contribute. 

Currently contributions to elastic scattering observables due to the spin-projected one-body
densities have only been calculated for light nuclei with $0^+$ ground states, and it was
found that this contribution is very small for nuclei with equal proton and neutron
numbers~\cite{Burrows:2020qvu,Baker:2021izp}. This is likely different for nuclei with
ground states of nonzero spin, which was explored for $^{10}$B polarization transfer observables
in Ref.~\cite{Cunningham:2011zz,Cunningham:2013lga}, where the authors assume a
nuclear structure which consists of  a core and valence nucleons. The work of Ref.~\cite{Vorabbi:2021kho} 
extends the standard leading order calculation to nonzero spin nuclei, however does not
consider the  inherent tensor contributions from the NN force in their formulation. This
leaves the importance of a consistent treatment of the NN force on elastic scattering from
nonzero spin nuclei still an open question.

The complete calculation of the leading-order effective interaction describing the scattering
of a proton from a nucleus in a $0^+$ ground state and which enters the integral
Eq.~(\ref{eq:2.11}) as driving term is given by
\begin{eqnarray}
  \label{eq:3.3}
  \lefteqn{\widehat{U}_{\mathrm{p}}(\bm{q},\bm{\mathcal{K}}_{NA},\epsilon) =} & &  \\
  & & \sum_{\alpha=\mathrm{n,p}} \int d^3{\mathcal{K}} \eta\left( \bm{q}, \bm{\mathcal{K}}, \bm{\mathcal{K}}_{NA} \right)
  A_{\mathrm{p}\alpha}\left( \bm{q}, \frac{1}{2}\left( \frac{A+1}{A}\bm{\mathcal{K}}_{NA} - \bm{\mathcal{K}} \right); \epsilon
  \right) \rho_\alpha^{K_s=0} \left(\bm{\mathcal{P}'}, \bm{\mathcal{P}}  \right) \cr
  &+&i (\bm{\sigma^{(0)}}\cdot\hat{\bm{n}}) \sum_{\alpha=\mathrm{n,p}} \int d^3{\mathcal{K}} \eta\left(\bm{q},
  \bm{\mathcal{K}}, \bm{\mathcal{K}}_{NA} \right)
  C_{\mathrm{p}\alpha}\left( \bm{q}, \frac{1}{2}\left( \frac{A+1}{A}\bm{\mathcal{K}}_{NA} - \bm{\mathcal{K}} \right);
  \epsilon
  \right) \rho_\alpha^{K_s=0} \left(\bm{\mathcal{P}'}, \bm{\mathcal{P}}  \right) \cr
  &+&i \sum_{\alpha=\mathrm{n,p}} \int d^3{\mathcal{K}} \eta\left( \bm{q}, \bm{\mathcal{K}}, \bm{\mathcal{K}}_{NA}
  \right) C_{\mathrm{p}\alpha} \left( \bm{q}, \frac{1}{2}\left( \frac{A+1}{A}\bm{\mathcal{K}}_{NA} - \bm{\mathcal{K}}
  \right); \epsilon \right) S_{n,\alpha} \left(\bm{\mathcal{P}'}, \bm{\mathcal{P}} \right) \cos \beta\cr
  &+&i (\bm{\sigma^{(0)}}\cdot\hat{\bm{n}}) \sum_{\alpha=\mathrm{n,p}} \int d^3{\mathcal{K}} \eta\left(\bm{q},
  \bm{\mathcal{K}}, \bm{\mathcal{K}}_{NA} \right)  (-i)
  M_{\mathrm{p}\alpha} \left( \bm{q}, \frac{1}{2}\left( \frac{A+1}{A}\bm{\mathcal{K}}_{NA} - \bm{\mathcal{K}}
  \right); \epsilon \right) S_{n,\alpha} \left(\bm{\mathcal{P}'}, \bm{\mathcal{P}}  \right) \cos \beta  . \nonumber
  \end{eqnarray}
 The term $\eta\left( \bm{q}, \bm{\mathcal{K}}, \bm{\mathcal{K}_{NA}} \right)$  is the M{\o}ller
  factor~\cite{CMoller} describing the transformation from the NN frame to the NA frame.
  The functions $A_{\mathrm{p}\alpha}$, $C_{\mathrm{p}\alpha}$, and $M_{\mathrm{p}\alpha}$ represent the NN interaction through Wolfenstein
amplitudes~\cite{wolfenstein-ashkin}. Since the incoming proton can interact
  with either a proton or a neutron in the nucleus, the index $\alpha$ indicates the
  neutron ($\mathrm{n}$) and proton ($\mathrm{p}$) contributions, which are calculated separately and then summed up.
 With respect to the nucleus, the operator $i (\bm{\sigma^{(0)}}\cdot \hat{\bm{n}})$ represents the spin-orbit
operator in momentum space with respect to the projectile. As such, Eq.~(\ref{eq:3.3}) exhibits the
  expected form of an interaction between a spin-$\frac{1}{2}$ projectile and a target nucleus in a $J=0$ state \cite{RodbergThaler}.
The momentum variables in the problem are given as
\begin{eqnarray}
\label{eq:3.4}
  \bm{q} &=& \bm{p'} - \bm{p} = \bm{k'} - \bm{k}, \\
  \bm{\mathcal{K}} &=& \frac{1}{2} \left(\bm{p'} + \bm{p}\right), \cr
  \bm{\mathcal{K}_{NA}} &=& \frac{A}{A+1}\left[\left(\bm{k'} + \bm{k}\right) +
        \frac{1}{2} \left(\bm{p'} + \bm{p}\right) \right], \cr
  \bm{\mathcal{P}}&=& \bm{\mathcal{K}}+\frac{A-1}{A}\frac{\bm{q}}{2},  \cr
  \bm{\mathcal{P'}}&=& \bm{\mathcal{K}}-\frac{A-1}{A}\frac{\bm{q}}{2} \nonumber  .
  \end{eqnarray}
The two quantities representing the structure of the nucleus are the scalar one-body density
  $\rho_\alpha^{K_s=0} \left(\bm{\mathcal{P}'}, \bm{\mathcal{P}}  \right)$ and the
  spin-projected momentum distribution $S_{n,\alpha} \left(\bm{\mathcal{P}'}, \bm{\mathcal{P}}
  \right)= \rho^{K_s=1}\left(\bm{\mathcal{P}'}, \bm{\mathcal{P}}\right)  \cdot \bm{\hat n}$. 
 Both distributions are nonlocal and translationally invariant. The reduced matrix elements entering the
one-body densities are obtained within the NCSM (SA-NCSM) in the center-of-mass frame of the nucleus. 
In order to employ them in calculating the leading-order effective NA interaction, this center-of-mass variable must
be removed. Within the framework of NCSM (SA-NCSM) the technique for obtaining nonlocal and translationally
invariant one-body densities is well
developed~\cite{Navratil:2004dp,Cockrell:2012vd,Mihaila:1998qr,Gennari:2017yez,Burrows:2018ggt,Navratil:2021mkp}.
  Lastly, the term $\cos \beta$ in Eq.~(\ref{eq:3.3}) results from projecting $\bm{\hat{n}}$ from the NN frame to the NA frame. For further details, see Ref.~\cite{Burrows:2020qvu}.


\section{Chiral Truncation uncertainties in the leading order optical potential}

With the emergence of nuclear forces based on chiral effective field theory (EFT), we are 
presented with an opportunity to study the nucleon-nucleus effective interaction as it develops 
order-by-order in a chiral EFT framework. Given the hierarchical nature of chiral EFT, we can 
combine these order-by-order results to reliably estimate truncation uncertainties associated 
with the higher chiral orders not included in the calculations. 
To this end, Refs.~\cite{Melendez:2017phj,Melendez:2019izc,Epelbaum:2019zqc}  first implemented 
uncertainty quantification for the cases of NN and Nd scattering by assuming a quantity $y(x)$ 
at a chiral order $k$ can be written as

\begin{eqnarray}
y_k(x) = y_{\mathrm{ref}}(x) \sum_{n=0}^k c_n(x) Q^n(x)
\label{eqn:yk}
\end{eqnarray}
where $y_{\mathrm{ref}}(x)$ is a reference value that sets the scale of the problem and also 
includes the dimensions of the quantity $y(x)$ of interest. By construction, the coefficients $c_n(x)$ are dimensionless and are expected to be of order unity. The remaining quantity $Q(x)$ is the expansion parameter associated with the chiral EFT. The expansion parameter is usually defined as 
\begin{eqnarray}
Q = \frac{1}{\Lambda_b}\max(M_{\pi}, p)
\end{eqnarray}
where $\Lambda_b$ is the breakdown scale of the EFT, $M_{\pi}$ is the pion mass, and $p$ is the relevant momentum for the problem. Various works~\cite{Melendez:2017phj,Melendez:2019izc,Epelbaum:2019zqc}
  have identified the relevant momentum in different ways, but keeping with Ref.~\cite{Baker:2021iqy}
 we choose the relevant momentum as the center-of-mass (c.m.) momentum in the nucleon-nucleus system
\begin{eqnarray}
p^2_{NA} = \frac{E_{\mathrm{lab}}A^2 m^2 (E_{\mathrm{lab}}+2m)}{m^2 (A+1)^2 + 2AmE_{\mathrm{lab}}}
\label{eqn:pNA}
\end{eqnarray}
where $E_{\mathrm{lab}}$ is the kinetic energy of the projectile in the laboratory frame, $A$ is the target nucleus's mass number, and $m$ is the mass of the nucleon. 

\begin{figure}
\begin{center}
\includegraphics[width=0.75\textwidth]{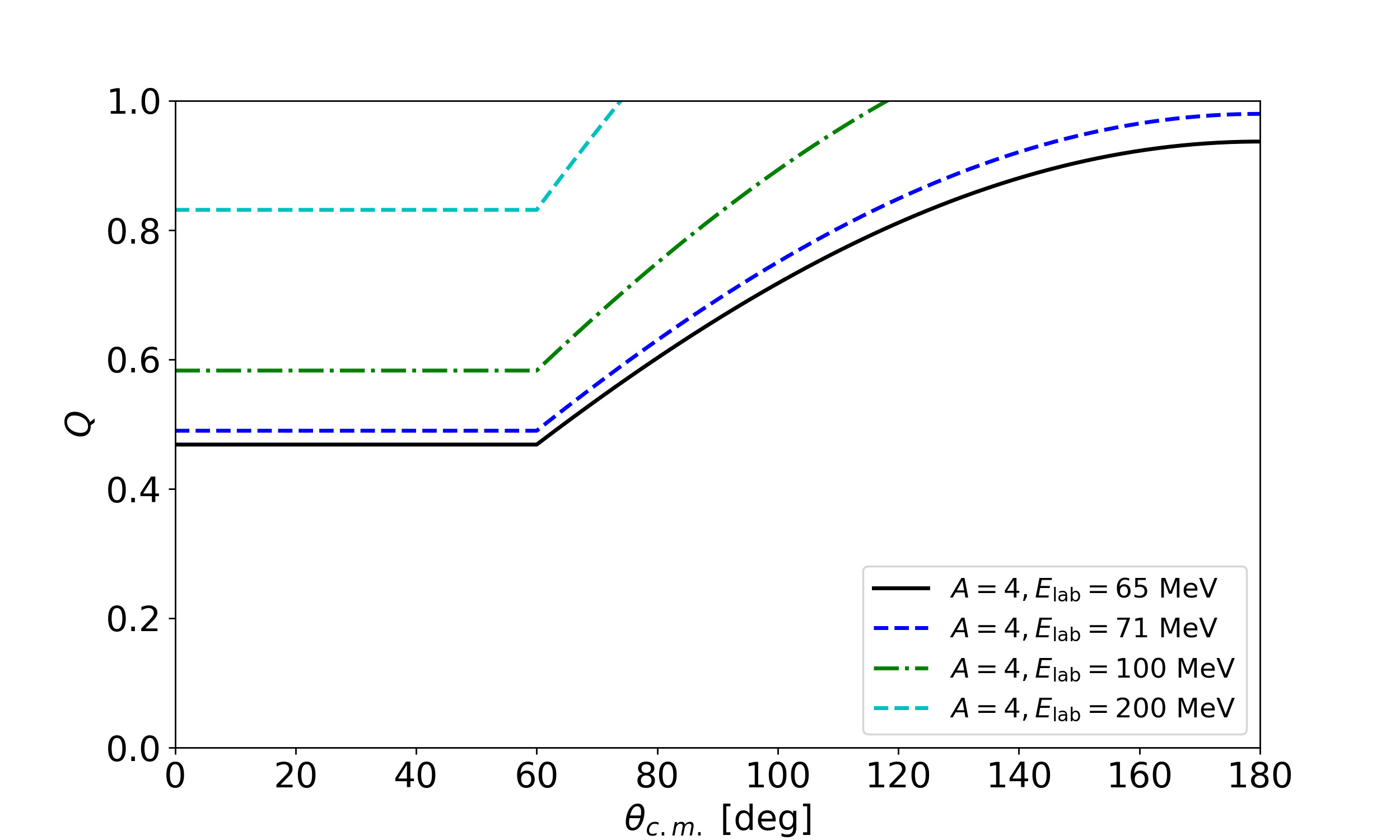}
\end{center}
\caption{The expansion parameter $Q$, defined by Eq.~(\ref{eqn:Q_with_q}) where $\Lambda_b=600$ MeV, as a function of the center-of-mass angle $\theta_{\mathrm{c.m.}}$ for a range of lab projectile energies $E_{\mathrm{lab}}$. In this case of nucleon-nucleus (NA) elastic scattering, the transition between when the expansion parameter is dominated by the center-of-mass momentum and the momentum transfer can easily be identified.}
\label{fig1}
\end{figure}

Previous scattering works \cite{Melendez:2019izc,Baker:2021iqy} have noted that various results indicate, when identifying the relevant momentum, the momentum transfer $q$ should also be considered. That is, the expansion parameter would be more appropriately defined as
\begin{eqnarray}
Q = \frac{1}{\Lambda_b}\max(M_{\pi}, p_{NA}, q)
\label{eqn:Q_with_q}
\end{eqnarray}
The momentum transfer in elastic scattering is defined as
\begin{eqnarray}
q = 2p_{NA} \sin \left (\frac{\theta_{\mathrm{c.m.}}}{2} \right)
\end{eqnarray}
where $\theta_{\mathrm{c.m.}}$ is the scattering angle in the c.m. frame. Notably, including the momentum transfer in Eq.~(\ref{eqn:Q_with_q}) makes the expansion parameter a function of $\theta_{\mathrm{c.m.}}$, even though the other momentum scales in Eq.~(\ref{eqn:Q_with_q}) are independent of the scattering angle. When considering observables such as the differential cross section or analyzing power that are functions of $\theta_{\mathrm{c.m.}}$, this implies the expansion parameter will be larger at backward angles than at forward angles. Furthermore, since the leading order of the spectator expansion is not applicable at low energies, we only consider scattering at lab energies of 65 MeV or higher. As a result, the chiral expansion parameter becomes $Q = \max(p_{NA}, q)/\Lambda_b$. This expansion parameter is shown in Fig.~\ref{fig1} for the case of $A=4$ and $\Lambda_b = 600$ MeV. Because of the factorization of the c.m. momentum, there is a universal scattering angle at which the momentum transfer $q$ begins to dominate the expansion parameter, regardless of the chosen $E_{\mathrm{lab}}$ or nucleus. We will exploit this behavior in later sections.

\subsection{Nuclear structure calculations}

Prior to our detailed study of truncation uncertainties of a chiral NN interaction in elastic NA
scattering observables we need to choose a specific chiral NN interaction. Here we want to focus on
the EKM chiral NN interaction~\cite{Epelbaum:2014sza,Epelbaum:2014efa} with a semi-local coordinate
space regulator of R~=~1~fm, which has a breakdown scale of $\Lambda_b$~=~600~MeV. 
This interaction gives a slightly better description of the ground state
energies in the upper $p$-shell than a similar, more recent interaction with a semi-local momentum space
regulator. For consistency with the leading-order optical we only use the NN potentials, omitting
three-nucleon forces, which appear at N2LO in the chiral expansion, both in the structure and the
scattering part of the calculations. Including three-nucleon forces consistently in both, the
structure and scattering calculations requires going beyond the leading-order optical potential, and
is beyond the scope of this work. Though initial attempts of incorporating three-nucleon forces as an
effective density-dependent NN force in the scattering part have been
presented~\cite{Vorabbi:2020cgf}, they can not yet be considered as systematic consideration of
three-nucleon forces in NA scattering. For similar reasons, we restrict most of our results to N2LO
since three-nucleon force contributions at N3LO and N4LO are significant~\cite{Binder:2015mbz}.

Next, the translationally-invariant one-body density
  needed for the scattering calculation can be obtained using the
  NCSM approach, in which the nuclear wavefunction is expanded in
  Slater determinants of harmonic oscillator basis
  functions~\cite{BarrettNV13}.  Ideally, one uses a sufficiently
  large basis to ensure convergence of this expansion, but in practice
  observables depend on both the many-body basis truncation,
  $N_{\mathrm{max}}$ (defined as the total number of harmonic
  oscillator quanta in the many-body system above the minimal
  configuration), and on the harmonic oscillator scale $\hbar\Omega$.
  In Table~\ref{Tab:NucStruct_alt} we give the ground state binding
  energies and point-proton radii of $^4$He, $^{12}$C, and $^{16}$O
  obtained with the EKM chiral NN
  potential~\cite{Epelbaum:2014sza,Epelbaum:2014efa} with a semi-local
  coordinate space regulator of R~=~1~fm (note that at N2LO we did not
  include any three-nucleon forces).

For $^4$He we can obtain nearly converged results for
  both the binding energy and the proton radius, and these results
  agree, to within their estimated numerical uncertainties (the first
  set of uncertainties in Table~\ref{Tab:NucStruct_alt}), with
  Yakubovsky calculations using the same NN
  potential~\cite{Binder:2015mbz}.  However, for larger nuclei such as
  $^{12}$C and $^{16}$O we are more limited in the $N_{\mathrm{max}}$
  values that can be reached on current computational
  resources.~\footnote{One commonly applies a Similarity
    Renormalization Group (SRG) transformation to the NN potential in
    order to improve the convergence of the many-body calculation.
    However, this leads to induced three-nucleon forces that are
    non-negligible; omitting those would lead to a strong dependence
    on the SRG parameter.  We therefore choose to not employ such a
    transformation here.  For the binding energies we use an
  exponential extrapolation to the complete basis, with associated
  uncertainties, see Ref.~\cite{Binder:2015mbz} for details.  Radii
  converge rather slowly in a harmonic oscillator basis, and they do
  not necessarily converge monotonically with increasing
  $N_{\mathrm{max}}$; furthermore, in the scattering calculations we
  use densities obtained at fixed values of the harmonic oscillator
  parameters $N_{\mathrm{max}}$ and $\hbar\Omega$.  We therefore
  simply give in Table~\ref{Tab:NucStruct_alt} our results for the
  point-proton radii of $^{12}$C and $^{16}$O at
  $N_{\mathrm{max}}=10$, averaged over the range $ 16 \le \hbar\Omega
  \le 28$~MeV (the same range as is used for the scattering calculations).
  The numerical uncertainty estimates for the radii listed in
  Table~\ref{Tab:NucStruct_alt} correspond to the spread over this
  $\hbar\Omega$ interval; this is a systematic uncertainty due to the
  Gaussian fall-off of harmonic oscillator basis functions, and is
  therefore strongly correlated for the different chiral orders.
  However, the trend of a significant increase in the radii going
  from LO to NLO, followed by a smaller increase going from NLO to
  N2LO, is robust, and correlates with the decrease in binding
  energies going from LO to NLO to N2LO.
  Note that we did not include any
  chiral EFT corrections to the $R^2$ operator; and the experimental
  point-proton radii are extracted from the charge radius measured in
  electron scattering experiments, using standard proton and neutron
  finite-size corrections, relativistic corrections, and
  meson-exchange corrections. }

\subsection{Pointwise truncation uncertainties}
To assess the relative size of chiral truncation uncertainties compared to other known uncertainties, e.g. the harmonic oscillator parameters $N_{\mathrm{max}}$ and $\hbar\Omega$, we employ a pointwise truncation procedure and study reaction observables that are not functional quantities, e.g. reaction cross sections at a specified laboratory energy. This pointwise approach was previously implemented in 
Refs.~\cite{Melendez:2019izc,Baker:2021iqy}  and it starts by assuming the expansion parameter $Q$ and reference scale $y_{\mathrm{ref}}$ are known. From there, we can apply Eq.~(\ref{eqn:yk}) to calculate the coefficients $c_n$, which are treated as independent draws from the same underlying distribution. The properties of this distribution can be learned from Bayesian techniques and the posterior distribution for the prediction can be readily calculated with its associated credible intervals. For more details, see Ref.~\cite{Melendez:2019izc}.

In order to estimate the chiral truncation
  uncertainties of the obtained ground state binding energies and
  radii, we apply the pointwise approach with $Q \approx 0.3$ as the
  effective expansion parameter, following Ref.~\cite{Binder:2018pgl}.
  These uncertainties are listed as the second set of uncertainties in
  Table~\ref{Tab:NucStruct_alt}, starting from NLO.  Here we see that
  for the energies, the chiral uncertainties are at least of the same
  order as the estimated numerical uncertainties; however, the uncertainties of
  the radii of $^{12}$C and $^{16}$O are clearly dominated by their systematic
  dependence on the basis parameter $\hbar\Omega$. 

\begin{figure}
\begin{center}
\begin{tabular}{c}
\includegraphics[width=0.65\textwidth]{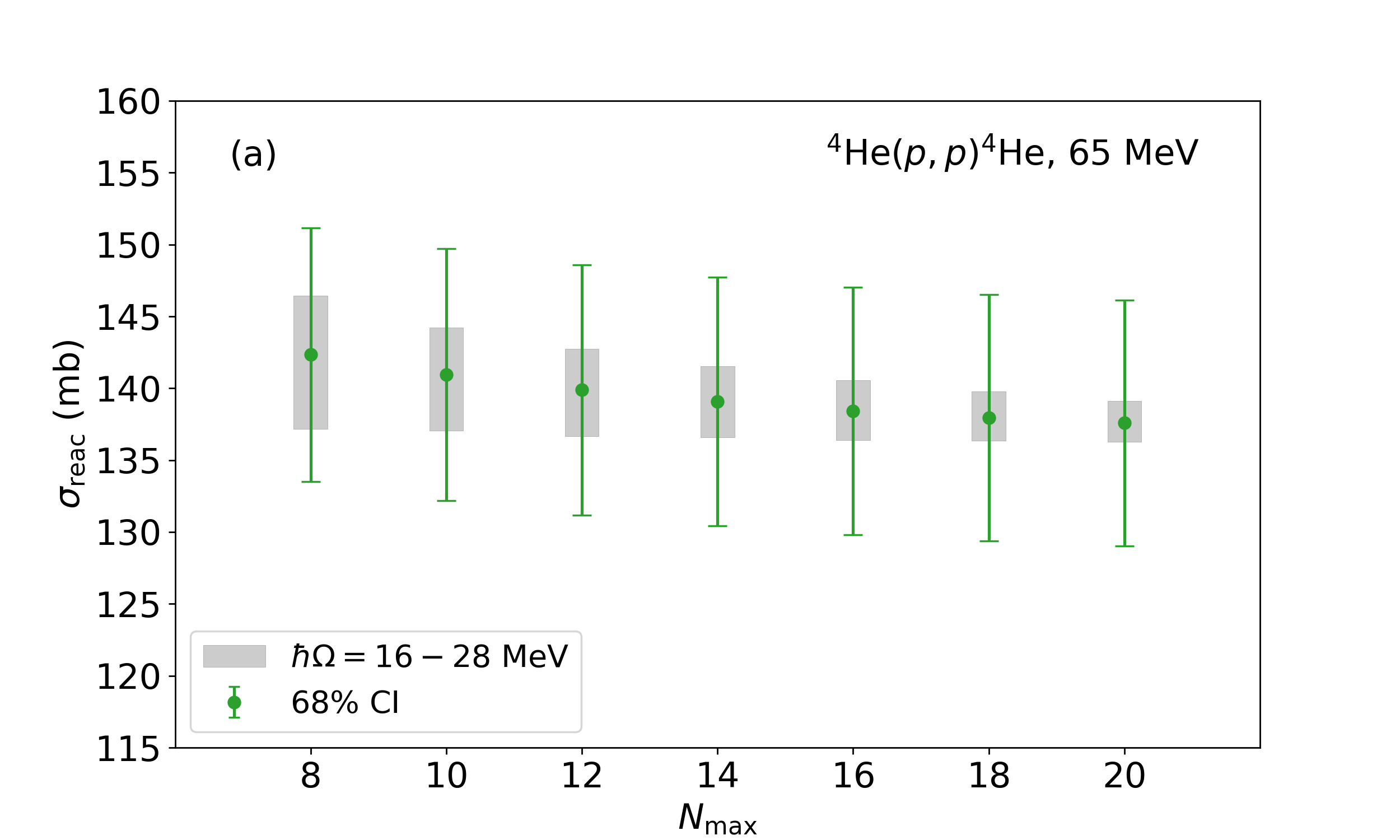} \\
\includegraphics[width=0.65\textwidth]{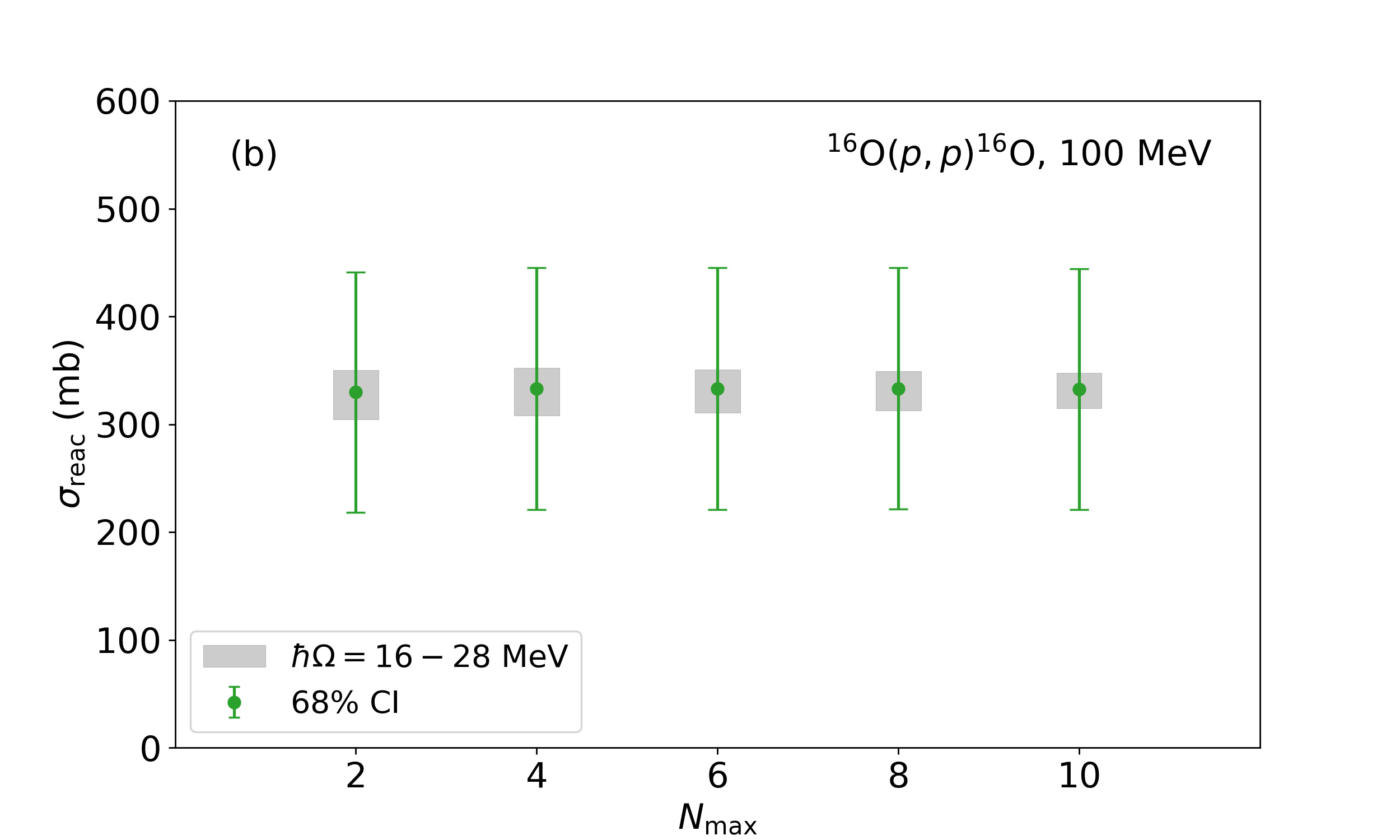}
\end{tabular}
\end{center}
	\caption{Reaction cross section for proton scattering on (a) $^4$He at 65 MeV and (b) $^{16}$O at 100 MeV, both at N2LO as a function of $N_{\mathrm{max}}$. The error bars show a 68\% credible interval (CI) from using a pointwise error estimation with the LO, NLO, and N2LO results. The shaded regions show variations with respect to the harmonic oscillator parameter $\hbar\Omega$. The values of the expansion parameters used were $Q=0.47$ for$^4$He at 65 MeV and $Q=0.69$ for $^{16}$O at 100 MeV. Note the different scales in (a) and (b).
	}
\label{fig2}
\end{figure}

To illustrate the pointwise approach
for scattering observables,
Fig.~\ref{fig2} shows the
reaction cross sections for proton scattering from $^4$He at 65~MeV
and $^{16}$O at 100~MeV.  For each case, the result is shown as a
function of $N_{\mathrm{max}}$, and variations with respect to
$\hbar\Omega$ are indicated.  While more obvious for the smaller
nucleus where the NCSM can better converge, in both cases the
uncertainty resulting from the chiral truncation remains larger than
the uncertainty arising from the many-body method.
To better illustrate this point, we present the
  reaction cross section for $^4$He with a scale starting from 115 mb
  and with a range of only 45 mb, while using the full range of 600 mb
  for $^{16}$O.
While larger model spaces will better converge the
NCSM results, smaller truncation uncertainties will only be achieved
by higher chiral orders, despite the noticeable
  dependence of the radii on the harmonic oscillator parameter
  $\hbar\Omega$, in particular for the heavier nuclei, in the current
  calculations.  Note however that even at N3LO we anticipate the
chiral truncation uncertainties will be larger than the indicated
variations with respect to the harmonic oscillator parameter
$\hbar\Omega$ due to the rather large value of the expansion parameter
$Q$ in the scattering calculation.

\subsection{Correlated truncation uncertainties}
For functional quantities $y(x)$ 
we employ a correlated approach that includes information at nearby values of $x$. This approach is better for observables such as a differential cross section, which we know does not vary wildly from values at nearby angles. It also starts from Eq.~(\ref{eqn:yk}) and treats the coefficients $c_n(x)$ as independent draws from an underlying Gaussian process. This Gaussian process encodes information about the correlation length $\ell$, and the qualities of the underlying distribution can be learned from the order-by-order results. This training is followed up by testing procedures which seek to confirm the Gaussian process has been appropriately fit to the available results, and if not, to diagnose potential issues. From a well-fit Gaussian process we can then extract truncation uncertainties for the functional quantities. For more details and applications, see Refs.~\cite{Melendez:2019izc,Baker:2021iqy}.

In the following examples, we examine proton scattering for $^4$He, $^{12}$C, and $^{16}$O at various projectile energies and compare to the available experimental data. In each case, we show the convergence with respect to chiral order and the resulting decrease in the size of the chiral truncation uncertainties, as well as discuss any associated physics insights. To avoid concerns about the expansion parameter increasing at larger angles, we mostly restrict our analysis to forward angles where we expect the expansion parameter to be independent of the scattering angle.

\begin{figure}
\begin{center}
\includegraphics[width=\textwidth]{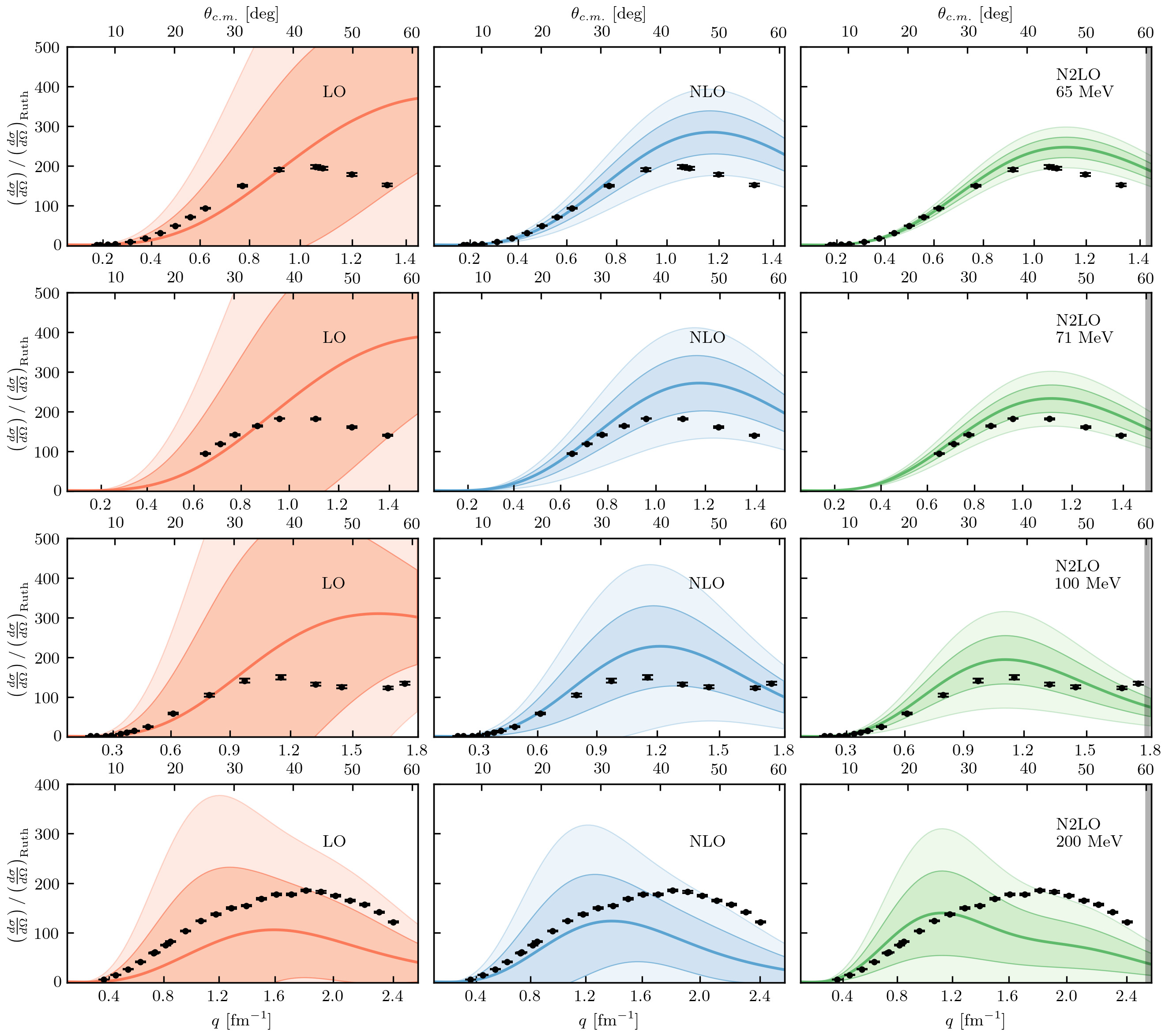}
\end{center}
\caption{Differential cross section divided by Rutherford for proton scattering on $^{4}$He at (first row) 65 MeV, (second row) 71 MeV, (third row) 100 MeV, and (fourth row) 200 MeV for LO (left column), NLO (middle column), and N2LO (right column) with corresponding $1\sigma$ (darker bands) and $2\sigma$ (lighter bands) error bands. Black dots are experimental data from Refs.~\cite{Imai:1979ihs} (65 MeV), \cite{Burzynski:1989zz} (71 MeV), \cite{Goldstein:1970dg} (100 MeV), and \cite{Moss:1979aw} (200 MeV).}
\label{fig3}
\end{figure}

For proton scattering on $^4$He, we see good agreement with experiment for the differential cross sections (Fig.~\ref{fig3}) at lower projectile energies. Below 100 MeV, most data points fall within the $2\sigma$ uncertainty band, and at 100 MeV a majority of the data points are within the $1\sigma$ band. At the highest energy of 200 MeV, the chosen interaction seems unable to reproduce the experimental data, though this is not uncommon for scattering from $^4$He.

\begin{figure}
\begin{center}
\includegraphics[width=\textwidth]{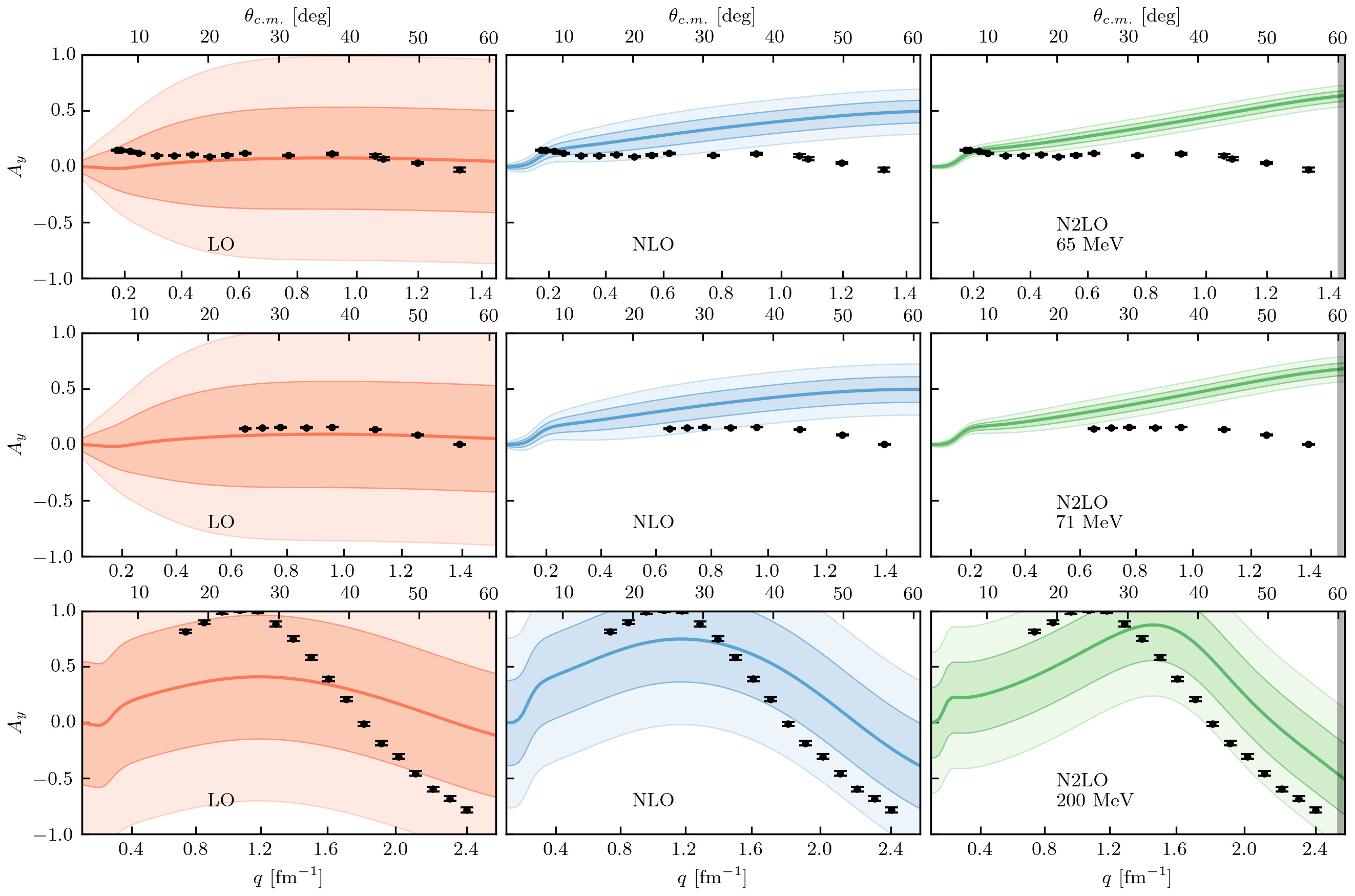}
\end{center}
\caption{Analyzing power for proton scattering on $^{4}$He at (first row) 65 MeV, (second row) 71 MeV, and (third row) 200 MeV) for LO (left column), NLO (middle column), and N2LO (right column) with corresponding $1\sigma$ (darker bands) and $2\sigma$ (lighter bands) error bands. Black dots are experimental data from Refs.~\cite{Imai:1979ihs} (65 MeV), \cite{Burzynski:1989zz} (71 MeV), and \cite{Moss:1979aw} (200 MeV).}
\label{fig4}
\end{figure}

The analyzing powers for proton scattering on $^4$He (Fig.~\ref{fig4}) is more complicated. For the lower energies of 65 and 71 MeV, the experimental data shows a near zero value, regardless of scattering angle. 
In the scattering of a spin-1/2 particle from a spin-0 nucleus, this indicates that there is no
spin-orbit force at play.
This behavior is only reproduced by the LO result, for which the chiral NN interaction only
contains the one-pion exchange and contact terms, which do not produce a spin-orbit force. At NLO the
two-pion exchange diagrams are responsible for reproducing the NN $p$-waves and thus provide a
spin-orbit force that leads to a non-zero value for the analyzing power in NA scattering. At N2LO
there are no new terms in the two-nucleon sector, and thus $A_y$ does not change its shape at that
chiral order. Therefore, one needs to conclude that in this case other physics which goes beyond the
leading order NA effective interaction may be needed to describe the analyzing power.

For the higher energy of 200 MeV, all of the experimental data points are within the $2\sigma$ uncertainty band, though there is a slight offset in the shape. In all cases, the analyzing power is more difficult to reproduce using this interaction, 
though other interactions have done better~\cite{Burrows:2017wqn,Burrows:2020qvu}

\begin{figure}
\begin{center}
\includegraphics[width=\textwidth]{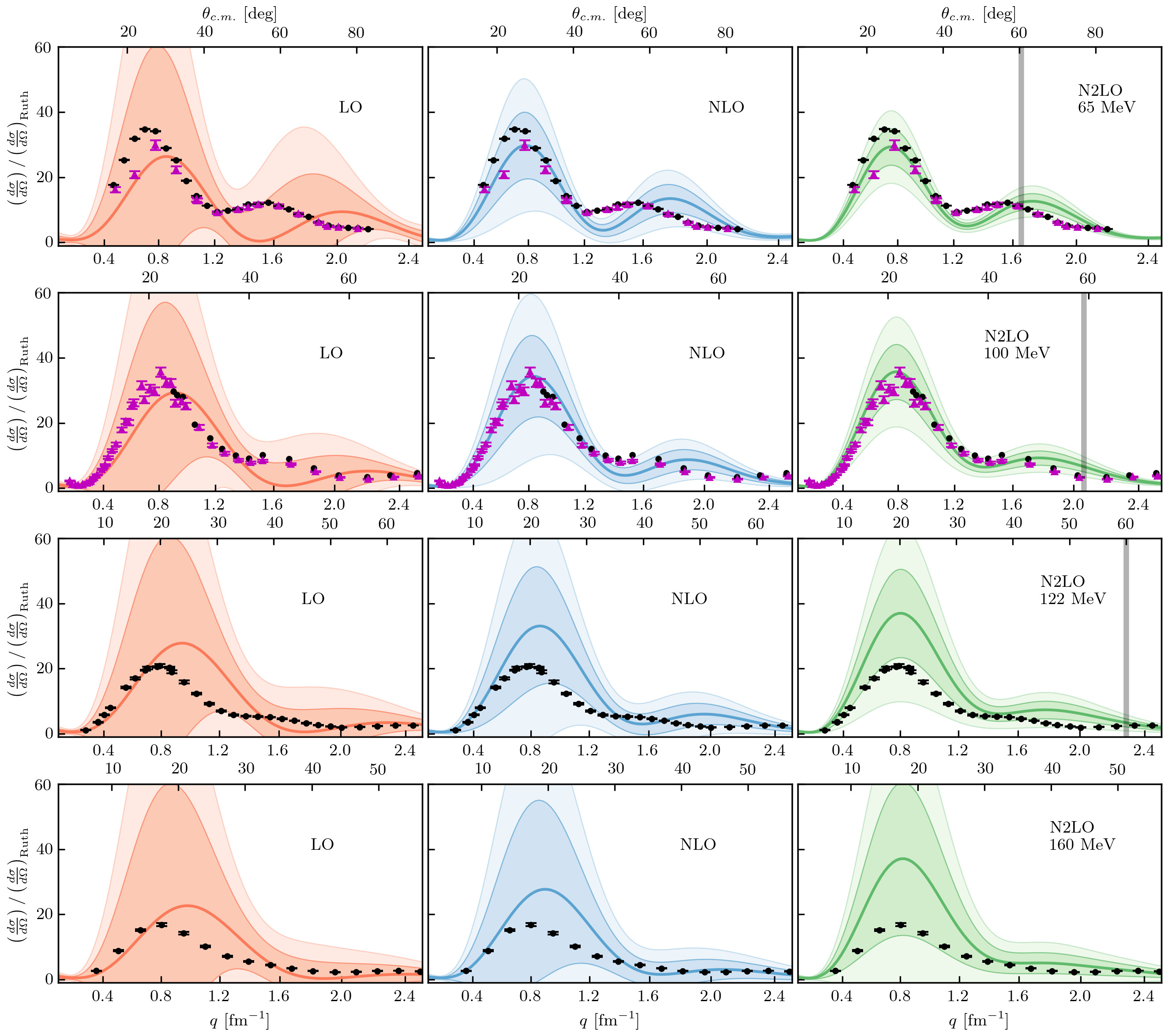}
\end{center}
\caption{
Differential cross section divided by Rutherford for proton scattering on $^{12}$C at (first row) 65 MeV, (second row) 100 MeV, (third row) 122 MeV, and (fourth row) 160 MeV for LO (left column), NLO (middle column), and N2LO (right column) with corresponding $1\sigma$ (darker bands) and $2\sigma$ (lighter bands) error bands. Black dots/purple triangles are experimental data from Refs.~\cite{Ieiri1987:sns} (65 MeV, black dots), \cite{Kato:1985koh} (65 MeV, purple triangles), \cite{Strauch:1956st} (96 MeV, purple triangles), \cite{Gerstein:1957gns} (99 MeV, black dots), and \cite{Meyer:1983kd} (122 MeV and 160 MeV).
Figure taken from Ref.\cite{Baker:2021iqy}.}
\label{fig5}
\end{figure}

For proton scattering from $^{12}$C, the differential cross sections (Fig.~\ref{fig5}) are  reliably 
reproduced by the central value of the N2LO calculations up to 100~MeV laboratory kinetic energy, 
and systematically over-predict at higher
energies. 
As the projectile energy increases, the expansion parameter increases and as a result uncertainty bands become larger. This is most noticeable at 160 MeV: the experimental data is within the $1\sigma$ band, 
but the size of that band, as well as the $2\sigma$ band, are so large that they are not practically useful.
The gray bars in the cross section panels for N2LO indicate the momentum transfer up to where we
expect the expansion parameter to be dominated by the c.m. momentum $p_{NA}$. Once the momentum
transfer exceeds the value given by the bar, the uncertainty is dominated by the momentum transfer
$q$, and is thus underrepresented by the method we use. Note that the vertical bar is at the same scattering angle $\theta_{\mathrm{c.m.}}$, but different momentum transfer $q$, as function of the
projectile energy since $p_{NA}$ is a function of the projectile energy as given in Eq.~(\ref{eqn:pNA}). 
Looking at the lower energies, the increasing agreement with experiment in the first peak and minimum
as higher orders in the chiral NN interaction are included gives the correct trend. Minima in the
differential cross section correlate with the size of the target nucleus. It is well well known
\cite{Binder:2018pgl}, and also evident from Table~\ref{Tab:NucStruct_alt},
that the nuclear binding energy calculated with the LO of the
chiral NN interaction is way too large and correspondingly the radius much too small. Only when
going
to NLO and N2LO the binding energy as well as the radius move into the vicinity of their experimental
values. This finding from structure calculations is corroborated by the calculations in
Fig.~\ref{fig5}, where with increasing chiral order the calculated first diffraction minimum moves
towards smaller momentum transfers indicating a larger nuclear size.

\begin{figure}
\begin{center}
\includegraphics[width=\textwidth]{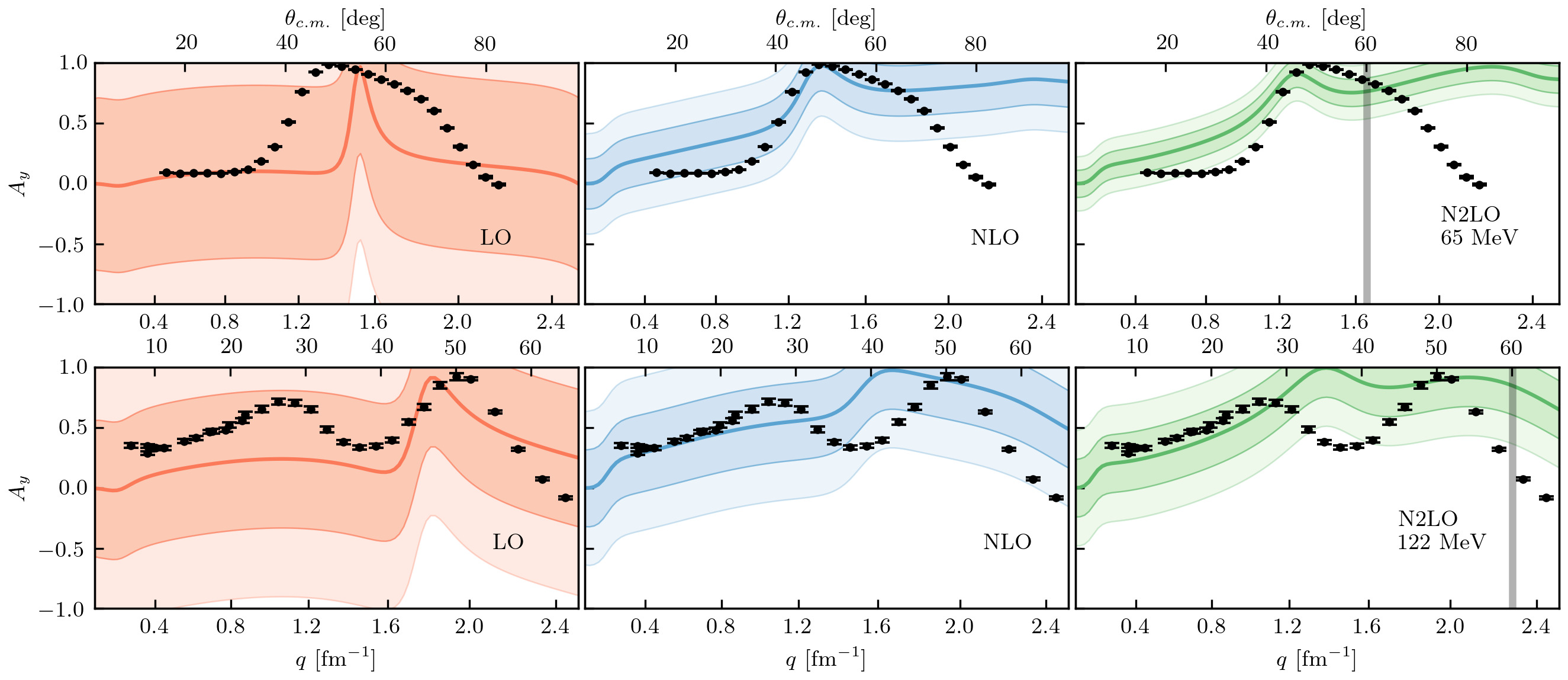}
\end{center}
\caption{Analyzing power for proton scattering on $^{12}$C at (first row) 65 MeV and (second row) 122 MeV for LO (left column), NLO (middle column), and N2LO (right column) with corresponding $1\sigma$ (darker bands) and $2\sigma$ (lighter bands) error bands. Black dots are experimental data from Refs.~\cite{Ieiri1987253} (65 MeV) and \cite{Meyer:1983kd} (122 MeV).
Figure taken from Ref.~\cite{Baker:2021iqy}.}
\label{fig6}
\end{figure}

The analyzing powers for proton scattering on $^{12}$C are at 65 MeV also almost zero for small
momentum transfers and rise at $q$~=~1.2~fm$^{-1}$ to its maximum value of +1. This is captured by the
NLO calculation where spin-contributions occur in the NN interaction (Fig.~\ref{fig6}).  
For 65 MeV, the experimental data is mostly within the $2\sigma$ band until approximately 
$\theta_{\mathrm{c.m.}}=60^{\circ}$, where we expect the expansion parameter to being increasing 
and the uncertainty bands to thus be underestimates. For 122 MeV, the very forward direction 
is inside the $1\sigma$ band, but the overall shape of the experimental data is not well captured 
by this interaction.

\begin{figure}
\begin{center}
\includegraphics[width=\textwidth]{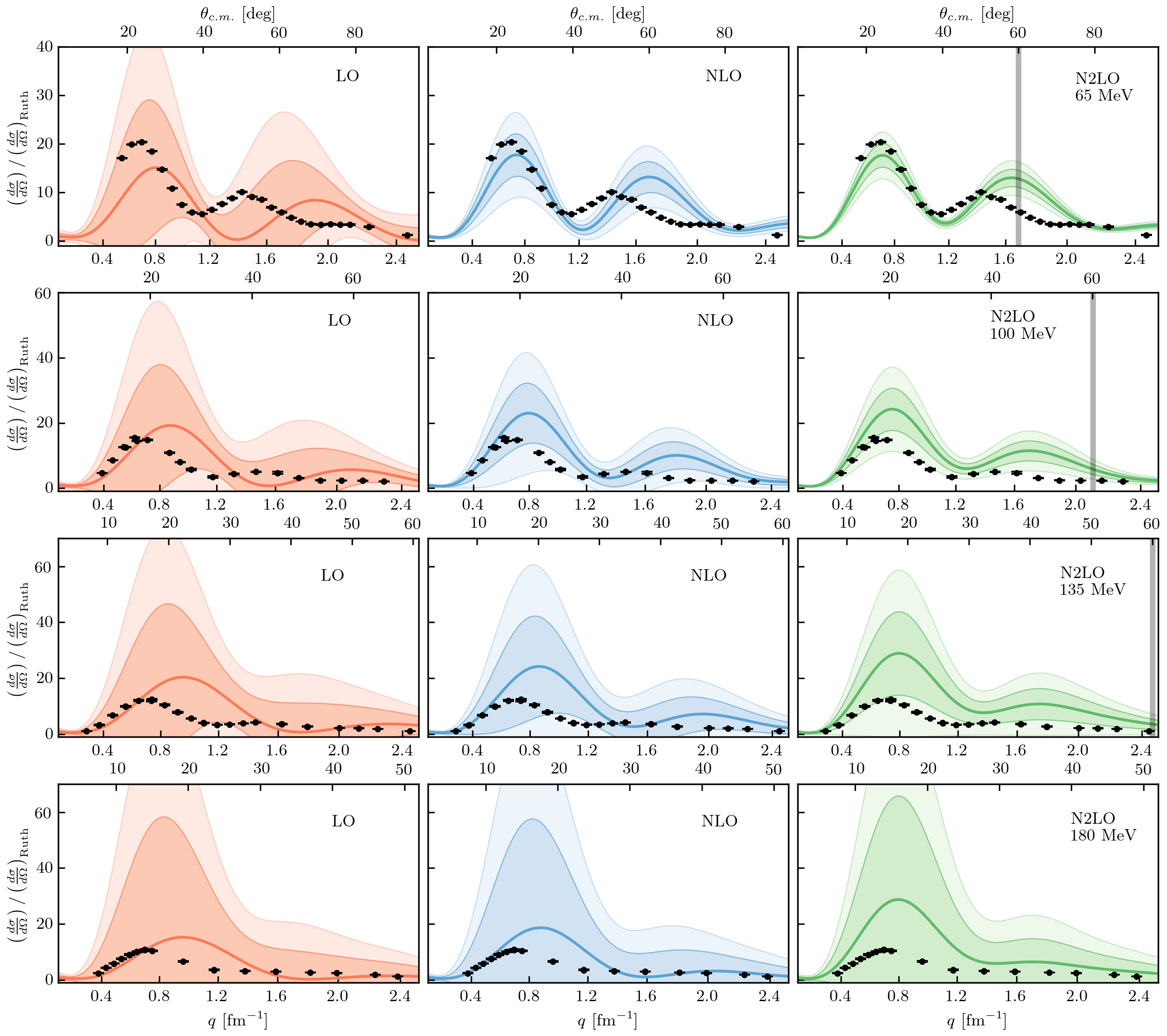}
\end{center}
\caption{Differential cross section divided by Rutherford for proton scattering on $^{16}$O at (first row) 65 MeV, (second row) 100 MeV, (third row) 135 MeV, and (fourth row) 180 MeV for LO (left column), NLO (middle column), and N2LO (right column) with corresponding $1\sigma$ (darker bands) and $2\sigma$ (lighter bands) error bands. Black dots are experimental data from Refs.~\cite{Sakaguchi:1979fpk} (65 MeV), \cite{Seifert:1990um} (100 MeV), \cite{Kelly:1989zza} (135 MeV), and \cite{Kelly:1990zza} (180 MeV).
Figure taken from Ref.~\cite{Baker:2021iqy}.}
\label{fig7}
\end{figure}

For proton scattering from $^{16}$O, the differential cross sections (Fig.~\ref{fig7}) are similar 
to the $^{12}$C case. Namely, the lower energies do reasonably well at describing the data within the $2\sigma$ bands, but as the projectile energy increases the uncertainty bands increase to unhelpful sizes. At the lowest energy of 65 MeV, we see a better and better reproduction of the first minimum in the differential cross section as the chiral order increases. Again, this first minimum is known to be related to the size of the nucleus, so this is an important feature to reproduce from both a structure, see Table~\ref{Tab:NucStruct_alt},
 and reaction perspective.

\begin{figure}
\begin{center}
\includegraphics[width=\textwidth]{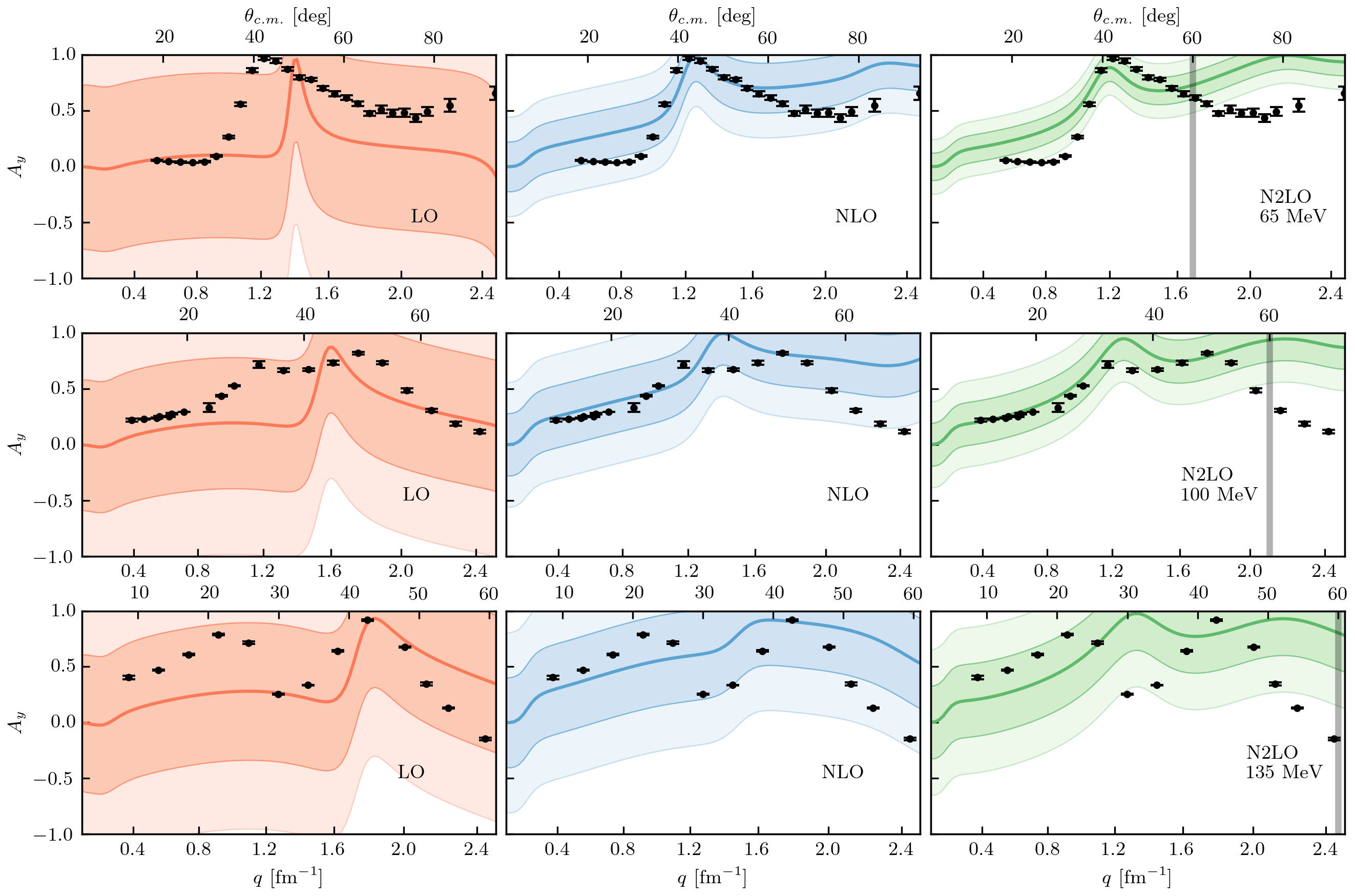}
\end{center}
\caption{Analyzing power for proton scattering on $^{16}$O at (first row) 65 MeV, (second row) 100 MeV, and (third row) 135 MeV for LO (left column), NLO (middle column), and N2LO (right column) with corresponding $1\sigma$ (darker bands) and $2\sigma$ (lighter bands) error bands. Black dots are experimental data from Refs.~\cite{Sakaguchi:1979fpk} (65 MeV), \cite{Seifert:1990um} (100 MeV), and \cite{Kelly:1989zza} (135 MeV).
Figure taken from Ref.~\cite{Baker:2021iqy}.}
\label{fig8}
\end{figure}

The analyzing powers for proton scattering on $^{16}$O (Fig.~\ref{fig8}) are again similar to the $^{12}$C case. At lower energies (65 and 100 MeV), we again see a good reproduction to within $1\sigma$ or $2\sigma$ of the forward direction data, but beyond $\theta_{\mathrm{c.m.}}=60^{\circ}$, the experimental data is outside the uncertainty bands. At the higher energy of 135 MeV, many of the experimental data are within the uncertainty bands but for a nucleus of this size, the expansion parameter has already increased such that the resulting uncertainty bands are unhelpfully large.

\begin{figure}
\begin{center}
\includegraphics[width=\textwidth]{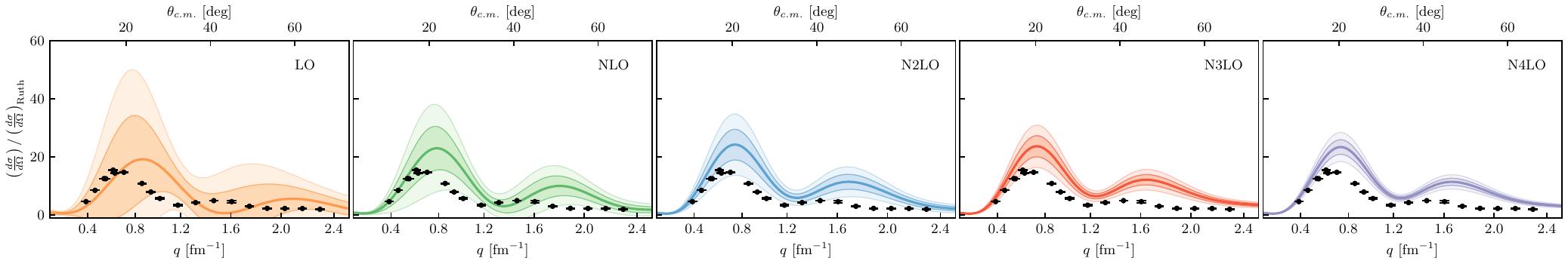}
\includegraphics[width=\textwidth]{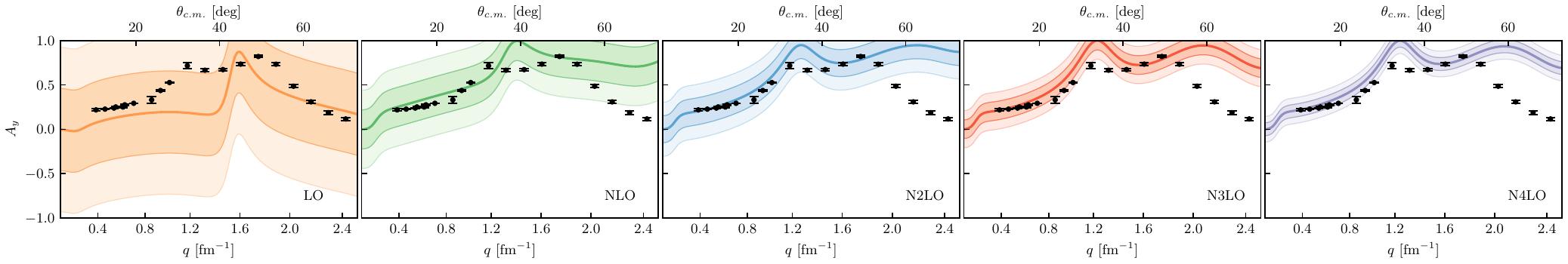}
\end{center}
\caption{Differential cross section divided by the Rutherford cross section (top) and
analyzing power (bottom) for proton scattering from $^{16}$O at 100~MeV. The first 3 columns are the
same as the second rows of Figs.~\ref{fig7} and \ref{fig8}. The additional two rightmost panels are
inconsistent calculations with use up to N2LO in the structure calculations and up to N3LO (fourth
column) or N4LO (fifth column) in the reaction calculation. Due to the inconsistency of the
calculation the uncertainty bands are not fully realistic. The data are the same as cited in
Figs.~\ref{fig7} and \ref{fig8}. 
}
\label{fig8b}
\end{figure}

As stated toward the beginning of the section we omit three-nucleon forces for consistency
with the leading-order optical potential which only treats two active nucleons. Those three-nucleon
forces already appear at N2LO in the chiral expansion, however, including them consistently in the
structure as well as reaction calculation requires going beyond the leading-order optical potential
and is beyond the scope of this work. For the sake of investigating truncation errors in the chiral NN
force, one may carry out inconsistent calculation in the sense that the structure part of the
calculation is kept fixed at N2LO, and in the reaction part higher orders in the NN force are used.
Proceeding in this fashion is sensible, since the scattering calculation is more sensitive
to the NN force compared to the structure calculation, provided this structure calculation gives a
reasonable description of the ground state one-body density. To show how the chiral truncation error
develops when higher chiral orders in the NN interaction are introduced, we show in Fig.~\ref{fig8b}
proton scattering from $^{16}$O at 100 MeV projectile energy, where the higher chiral orders are only
employed in the scattering part through the corresponding Wolfenstein amplitudes. In both, the
differential cross section as well as the analyzing power the two most right panels depicting the
inconsistent calculation show that the uncertainty bands become smaller when higher chiral orders in
the NN interaction are included. However, these uncertainty bands are not necessarily realistic due to
missing higher-body effects, which include higher orders in the chiral force as well as higher orders
in the multiple scattering expansion. Therefore, we can not draw firm conclusions from the fact that
data are outside the uncertainty estimates. Nevertheless, it is obvious that the decrease in the uncertainties in the chiral truncation is rather slow due to the large expansion parameter. 
Furthermore, the medians of the calculations shown in Figs.~\ref{fig7} and~\ref{fig8} 
do not change when higher chiral orders are considered in Fig.~\ref{fig8b}, which further indicates 
that the smaller error bands of the higher order chiral truncations may be artificial.

\subsection{Analysis of Posteriors}
Even while restricting our analysis to a region where we expect the expansion parameter to be constant, we can still observe effects on the uncertainty bands if the expansion parameter is large, as noted in many of the results at larger projectile energies. In fact, this behavior will place limits on the size of nucleus that can be considered with this approach, since $p_{NA}$ as defined by Eq.~(\ref{eqn:pNA}) will continue to increase as $A$ increases, yielding $Q>1$ eventually. While this situation is not ideal, we nonetheless find support for it in our analysis after examining the posteriors for $Q$, in accordance with Ref.~\cite{Baker:2021iqy,Melendez:2019izc}. 

\begin{figure}
\begin{center}
\begin{tabular}{c}
\includegraphics[width=0.4\textwidth]{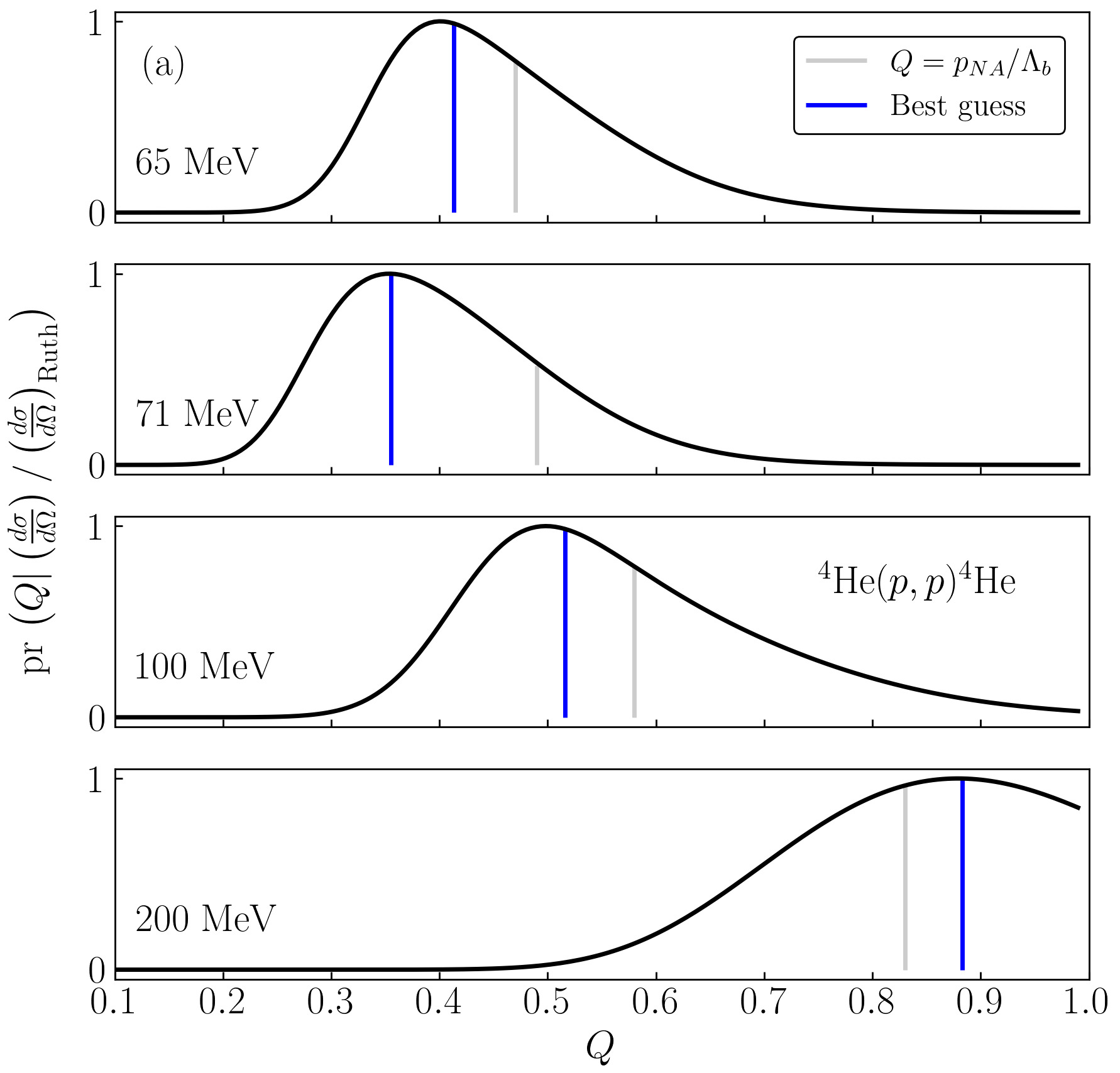}	\\
\includegraphics[width=0.4\textwidth]{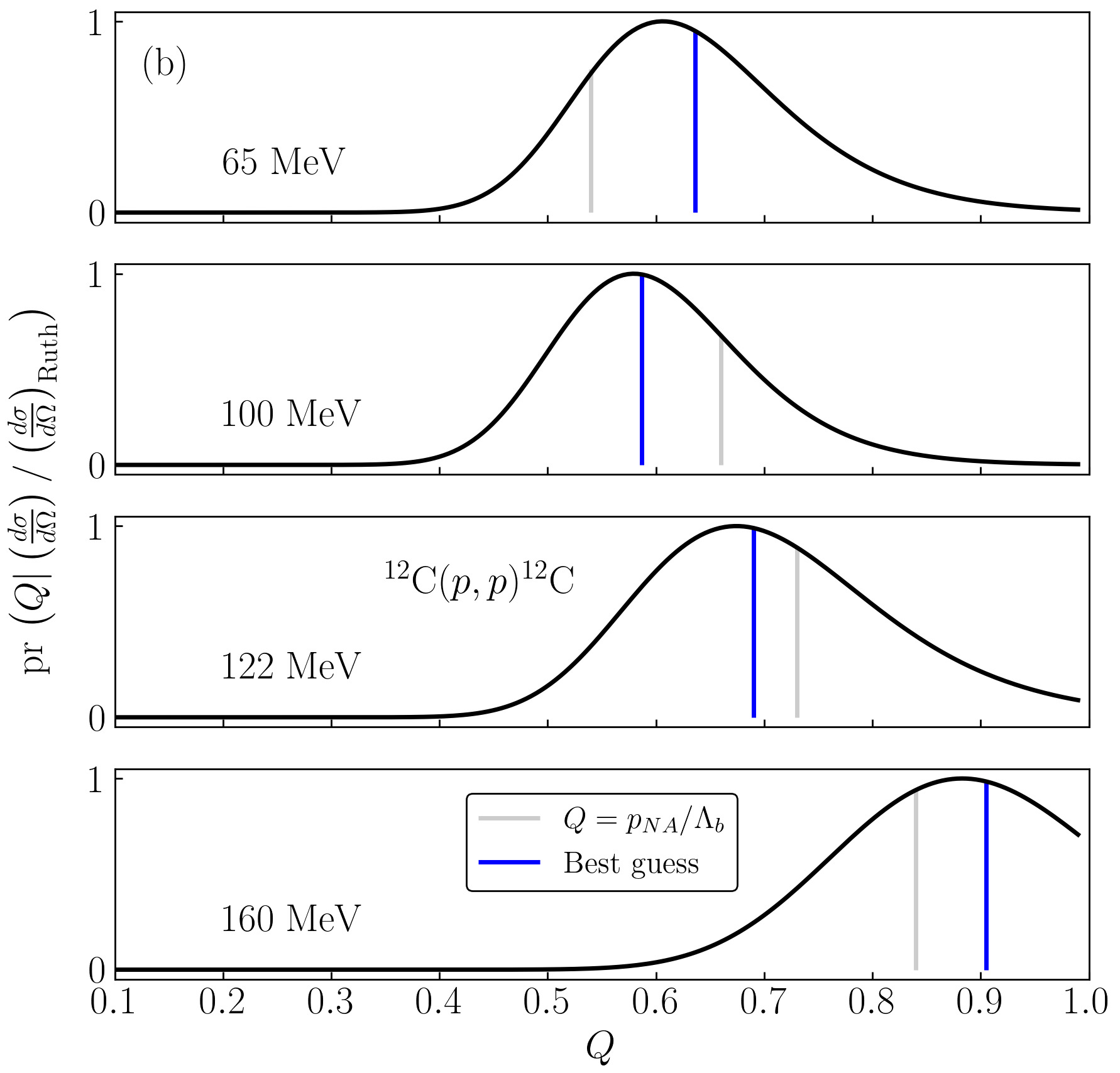}	\\
\includegraphics[width=0.4\textwidth]{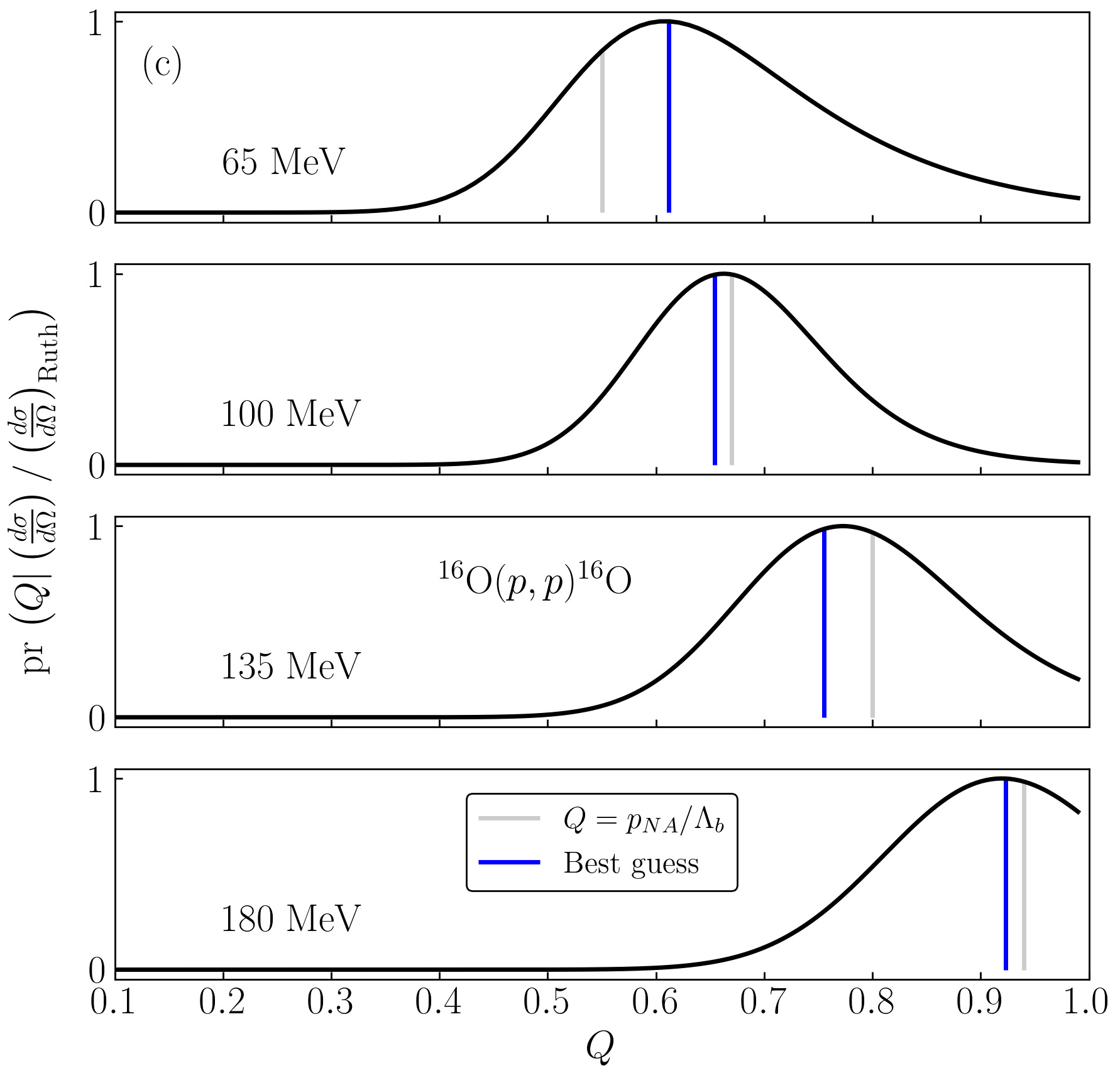}	
\end{tabular}
\end{center}
\caption{Posterior plots for the expansion parameter $Q$ given the differential cross sections for proton scattering on (a) $^4$He, (b) $^{12}$C, and (c) $^{16}$O.}
\label{fig9}
\end{figure}

In Fig.~\ref{fig9}, we calculated posteriors for the differential cross sections in proton scattering from $^4$He, $^{12}$C, and $^{16}$O at the energies previously discussed. From these, we can extract a single best guess for the value of $Q$ based on the order-by-order calculations and compare that to the expectation for $Q$ based on Eq.~(\ref{eqn:Q_with_q}). For $^{16}$O, the largest nucleus considered, we see generally good agreement between the expected value of $Q$ and the best guess value from the posteriors (Fig.\ref{fig9}c). However, as the nucleus decreases in size and as the laboratory energy decreases, some differences begin to emerge between the two values. In Fig.~\ref{fig9}b for $^{12}$C, the comparisons are roughly similar to the $^{16}$O case, but for the $^4$He analysis (Fig.~\ref{fig9}a), the differences are more pronounced, especially for the lower laboratory energies. A similar analysis of neutron scattering on $^{12}$C did not show any significant differences between the two values \cite{Baker:2021iqy}, which implies $^4$He may be the outlier in this approach. This analysis may imply scattering from $^4$He with projectiles at lower energies could be analyzed with a smaller expansion parameter $Q$, though the higher energy results still favor the larger expansion parameter. As the smallest nucleus considered here, it may also point to the few-body character of $^4$He, which has not historically been well captured in an optical potential approach.


\section{Outlook}

Procedures that quantify the theoretical uncertainties associated with the underlying chiral EFT NN
interaction are by now well established for the NN and nucleon-deuteron systems as well as nuclear
structure calculations, while the systematic study of chiral truncation uncertainty is not as widely
used in {\it ab initio} effective interaction employed to describe the scattering of protons or
neutrons from nuclei.  Contributing factors for this relatively slow development include that when
considering a multiple scattering approach to deriving this effective NA interaction in an {\it ab
initio} fashion only recent progress in calculating the leading-order term in the multiple scattering
approach has allowed to treat the NN interaction on the same footing in the structure and reaction
part~\cite{Burrows:2020qvu} by considering the spin of the struck target nucleon. Though 
calculations showed that the latter does not contribute significantly to observables when considering
scattering from nuclei with a $0^+$ ground state, one nevertheless needs a consistent {\it ab initio}
implementation of the leading-order term of the effective NA interaction in order to study
the theoretical uncertainties imprinted on NA observables by the chiral EFT NN interaction. 

In this work we carry out a systematic study of chiral truncation uncertainties of the EKM chiral
interaction on the {\it ab initio} effective NA interaction calculated in leading order of the
spectator expansion for $^4$He, $^{12}$C, and $^{16}$O. We find that this interaction allows for a
good description of experiment at energies around 100~MeV projectile kinetic energy and slightly
lower, provided we focus on regions of momentum transfer where the analysis of the EFT truncation
uncertainty is valid. When considering the lower energy of 65~MeV, the agreement with data starts to
deteriorate. This is an indication that errors other than the truncation error in the chiral
interaction should come into play, specifically errors that result from the spectator expansion
itself. Theoretical consideration of the next-to-leading-order term in the spectator expansion are
described in some detail in this work in order to lay out necessary theoretical and computational 
developments for this nontrivial endeavor. At at the next-to-leading order
 three-nucleon forces will naturally enter the effective interaction. 
At present this step has only been attempted in approximative
fashions, namely by approximating the next-to-leading order in the propagator expansion via a nuclear mean 
field force~\cite{Chinn:1995qn} or by introducing an effective, density dependent NN potential in the scattering part
of the calculation~\cite{Vorabbi:2020cgf}. Since we are not considering next-to-leading order terms in the spectator
expansion, we restrict our analysis to N2LO in the
chiral interaction and only consider two-nucleon forces. In this case the choice of the EKM interaction with
a semi-local coordinate space regulator of 1.0~fm is advantageous~\cite{Maris:2020qne}, 
since this specific interaction
gives a slightly better description of the ground state energies in the upper $p$-shell compared to
other more recent chiral EFT interactions when using two-nucleon interactions only.

In our study the chiral truncation errors at energies larger than 100~MeV increase considerably and
the agreement with experiment deteriorates. The increase in the chiral truncation error can simply be traced 
back to the expansion parameter in our approach is getting too large.  The deterioration of the agreement with
experiment when going to higher energies is more difficult to answer.  
One conclusion may be that the specific EKM chiral
interaction employed here in using the leading-order in the spectator expansion is not well suited to describe
proton-nucleus scattering observables for $^4$He, $^{12}$C, and $^{16}$O at higher energies.
For the chiral NN interaction from Ref.~\cite{Ekstrom13} this is not the case as shown in
Refs.~\cite{Burrows:2020qvu,Burrows:2018ggt}. Therefore one will have to investigate what features of a chiral NN
interaction are most relevant for a description of NA scattering observables for light nuclei.  

To put this in perspective, let us reconsider the basic ideas of the spectator expansion. By design, the leading-order
term should be dominant at energies 150 MeV projectile kinetic energy and higher, since the reaction time
of the projectile with nucleons inside the nucleus is short, and thus an `impulse approximation' is in
general very good.  However, we do not want 
to consider here projectile
energies larger than 400 MeV, where a relativistic treatment e.g. via the Dirac equation may be 
 preferred~\cite{Cooper:1993nx,Hynes:1985kp}. Thus at energies around 200 MeV the leading order term by design
should give a reasonably good description of NA scattering data. This has been the case in the microscopic
calculations of the 1990s (see e.g.
\cite{Crespo:1992zz,Crespo:1990zzb,Elster:1996xh,Elster:1989en,Arellano:1990xu,Arellano:1990zz}) and a set of
recent calculations with specific chiral NN interactions~\cite{Burrows:2018ggt,Burrows:2020qvu,Vorabbi:2020cgf}. 
Attempts to go beyond the leading order by incorporating 3NFs in a density dependent fashion into the many-body
propagator~\cite{Vorabbi:2020cgf} indicate that effects at 200 MeV are only visible at higher momentum transfer. 
In a similar fashion,
investigations going beyond the leading order term in Ref.~\cite{Chinn:1995qn} indicate that those effects become
important at around 100 MeV and at higher momentum transfers. Thus, if the 3NFs inherent in the chiral expansion
are needed to influence calculations with chiral NN forces in the leading order of the spectator expansion at
higher energies, then a new look at the interplay between NN and 3NFs in the leading-order spectator expansion
must be developed.

\begin{acknowledgments}
R.B.~B. and Ch.~E. gratefully acknowledge fruitful discussions with R.J.~Furnstahl and D.R.~Phillips  about
quantifying truncation errors in EFTs. Ch.~E. acknowledges useful discussions with A.~Nogga about the
LENPIC chiral NN interactions. \\
This work was performed in part under the auspices of the U.~S.~Department of Energy under contract
Nos.~DE-FG02-93ER40756, DE-SC0018223 and DE-SC0023495, and by the U.~S.~NSF (PHY-1913728). The
numerical
computations benefited from computing resources provided by the Louisiana Optical Network Initiative
and HPC resources provided by LSU, together with resources of the National Energy Research Scientific
Computing Center, a U.~S.~DOE Office of Science User Facility located at Lawrence Berkeley National
Laboratory, operated under contract No.~DE-AC02-05CH11231.

\end{acknowledgments}

\bibliography{all}


%
%
\section*{Tables}


\begin{table}[h]
  \begin{tabular}{c|D{.}{.}{12}|D{.}{.}{12}|D{.}{.}{12}}
    \\
    & \multicolumn{1}{c|}{$^4$He}
    & \multicolumn{1}{c|}{$^{12}$C}             
    & \multicolumn{1}{c}{$^{16}$O}
    \\
    \hline\\[-9pt]
    &
    \multicolumn{3}{c}{Binding energy (MeV)}
    \\
    \hline\\[-8pt]
    LO      &  45.45(0.01)      & 137. (1.)     &  224. (2.)     \\
    NLO     &  28.53(0.01)(3.5) &  97. (3.)(9.) &  156. (5.)(14.) \\
    N$^2$LO &  28.11(0.01)(0.9) &  94. (4.)(3.) &  149. (5.)(4.)  \\[4pt]
    expt    &  28.30            &  92.16        &  127.62 \\
    \hline\\[-9pt]
    &
    \multicolumn{3}{c}{Point-proton radius (fm)}
    \\
    \hline\\[-8pt]
    LO      &  1.08(0.02)       & 1.85(0.17)       & 1.8(0.2)        \\
    NLO     &  1.40(0.02)(0.08) & 2.04(0.16)(0.09) & 2.05(0.16)(0.10) \\
    N$^2$LO &  1.42(0.02)(0.02) & 2.12(0.15)(0.03) & 2.11(0.15)(0.03) \\[4pt]
    expt    &  1.46             & 2.32             & 2.58  \\
    \hline
  \end{tabular}
  \caption{\label{Tab:NucStruct_alt}    
    Ground state binding energies (top) and point-proton RMS radii (bottom)
    of $^4$He, $^{12}$C, and $^{16}$O with LO, NLO, and N$^2$LO
    LENPIC SCS NN potentials.  Both our estimated numerical uncertainties
    (first set of uncertainties) and chiral truncation uncertainty estimates
    (second set of uncertainties, not evaluated for LO) are given.  }
\label{table1}
\end{table}

\clearpage

\end{document}